\documentclass[conference]{IEEEtran}
\usepackage{cite}
\usepackage{amsmath}
\usepackage{amssymb}
\usepackage{amsfonts}
\usepackage{mathtools}
\usepackage{bbold}
\usepackage{soul}
\usepackage{booktabs}
\usepackage{float}
\usepackage{graphicx}
\usepackage{textcomp}
\usepackage{xcolor}
\usepackage{xspace}
\usepackage{mathrsfs}
\usepackage{hyperref}
\usepackage{url}
\usepackage[inline]{enumitem}
\usepackage{tikz}
\usetikzlibrary{shapes}
\usepackage{amsthm}
\usepackage{algorithm,algpseudocode}
\usepackage{dsfont}
\usepackage{subfig}
\usepackage{mwe}
\usepackage{wrapfig}
\usepackage{lipsum}
\usepackage{framed}
\usepackage{svg}
\usepackage{multirow}
\usepackage{soul}
\usepackage{wasysym}

\newcommand{\new}[1]{\textcolor{black}{#1}}

\newcommand{\ie}{\textit{i.e.,}\@\xspace}
\newcommand{\eg}{\textit{e.g.,}\@\xspace}
\newcommand{\etal}{\textit{et al.}\@\xspace}
\newcommand{\E}{\ensuremath{\mathbb{E}}}
\newcommand{\SISA}{\textbf{SISA} training\xspace}

\theoremstyle{definition}
\newtheorem{definition}{Definition}[section]

\newcommand\blfootnote[1]{%
  \begingroup
  \renewcommand\thefootnote{}\footnote{#1}%
  \addtocounter{footnote}{-1}%
  \endgroup
}

\usepackage{pstricks}
\pdfoutput=1
\begin{document}

\title{\vspace*{-0.5in}
{{\normalsize \rm In 42\textsuperscript{nd} {\em IEEE Symposium of Security and Privacy}\hrule}}
\vspace*{0.4in}Machine Unlearning}

\author{Lucas Bourtoule\text{*}\IEEEauthorrefmark{3}\IEEEauthorrefmark{4}, Varun Chandrasekaran\text{*}\IEEEauthorrefmark{2}, Christopher A. Choquette-Choo\text{*}\IEEEauthorrefmark{3}\IEEEauthorrefmark{4}, Hengrui Jia\text{*}\IEEEauthorrefmark{3}\IEEEauthorrefmark{4},\\ Adelin Travers\text{*}\IEEEauthorrefmark{3}\IEEEauthorrefmark{4}, Baiwu Zhang\text{*}\IEEEauthorrefmark{3}\IEEEauthorrefmark{4}, David Lie\IEEEauthorrefmark{3}, Nicolas Papernot\IEEEauthorrefmark{3}\IEEEauthorrefmark{4}\vspace*{0.15cm} \\ 
University of Toronto\IEEEauthorrefmark{3}, Vector Institute\IEEEauthorrefmark{4}, University of Wisconsin-Madison\IEEEauthorrefmark{2} 
}

\maketitle
\thispagestyle{plain}
\pagestyle{plain}

\begin{abstract}

Once users have shared their data online, it is generally difficult for them to 
revoke access and ask for the data to be deleted. Machine learning (ML)
exacerbates this problem because any model trained with said data
may have memorized it, putting users at risk of a successful privacy attack exposing their information.
Yet, having models unlearn is notoriously difficult.

We introduce \SISA, a framework that expedites the unlearning process by strategically limiting the influence of a data point in the training procedure. While our framework is applicable to any learning algorithm, it is designed to achieve the largest improvements for stateful algorithms like stochastic gradient descent for deep neural networks. \SISA reduces the computational overhead 
associated with unlearning, even in the worst-case setting where unlearning
requests are made uniformly across the training set. In some cases, the service provider may
have a prior on the distribution of unlearning requests that will be issued by users. 
We may take this prior into account to partition and order data 
accordingly, and further decrease overhead from unlearning.

Our evaluation spans several datasets from different domains, with corresponding motivations for unlearning. Under no distributional assumptions, for simple learning tasks, \new{we observe  that \SISA improves time to unlearn points from the Purchase dataset by $4.63\times$, and $2.45\times$} for the SVHN dataset, over retraining from scratch. \SISA also provides a speed-up of $1.36\times$ in retraining for complex learning tasks such as ImageNet classification; aided by transfer learning, this results in a small degradation in accuracy. Our work contributes to practical data governance in machine unlearning.

\end{abstract}

\blfootnote{\text{*}All student authors contributed equally and are ordered alphabetically.}
\section{Introduction}

Many applications of machine learning (ML) involve analyzing data that is collected from individuals. This data is often sensitive in nature and could include information like medical records~\cite{liu2017detecting} or personal emails~\cite{chen2019gmail}. Morever, data pipelines are often not static~\cite{he2014practical}: new data is collected regularly and incrementally used to further refine existing models following the online learning paradigm~\cite{shalev2012online}. 

Conversely, data may also need to be deleted. Recently introduced legislation, such as the General Data Protection Regulation (GDPR) in the European Union~\cite{mantelero2013eu}, the California Consumer Privacy Act~\cite{ccpa} in the United States, and PIPEDA privacy legislation in Canada~\cite{canada_law} include provisions that require the so-called \textit{right to be forgotten}~\cite{shastri2019seven}.  This requirement, which has been one of the most controversial in the GDPR,  mandates that companies take {\em reasonable steps} to achieve {\em the erasure of personal data concerning [the individual]}~\cite{gdpr}.  The unprecedented scale at which ML is being applied on personal data motivates us to examine how this right to be forgotten can be efficiently implemented for ML systems.

    Because ML models potentially memorize training data~\cite{fredrikson2015model,carlini2018secret}, it is important to unlearn what they have learned from data that is to be deleted. This problem is tangential to privacy-preserving ML---enforcing $\varepsilon$-differential privacy~\cite{dwork2014algorithmic} \new{with  $\varepsilon \neq 0$ does not alleviate the need for an unlearning mechanism.} Indeed, while algorithms which are differentially private guarantee a bound on how much individual training points contribute to the model and ensure that this contribution remains small~\cite{chaudhuri2011differentially,abadi2016deep}, there remains \textit{a non-zero} contribution from each point. If this was not the case, the model would not be able to learn at all (see \S~\ref{sec:definition}). In contrast, forgetting requires that \textit{a particular} training point have {\em zero} contribution to the model, which is orthogonal to the guarantee provided by differential privacy.

Having models forget necessitates knowledge of exactly how individual training points contributed to model parameter updates. Prior work showed this is possible when the learning algorithm queries data in an order that is decided prior to the start of learning~\cite{cao_towards_2015} \new{\ie in the statistical query (SQ) learning setting~\cite{kearns1998efficient}}. When the dataset is instead queried adaptively, \ie a given query depends on any queries made in the past, convergence of the approach is no longer guaranteed. In the adaptive setting, the divergence induced by this approach is bounded only for models which require a small number of iterations for learning. \new{While it is true that any algorithm in the PAC setting can be converted to its equivalent in the SQ learning setting~\cite{kearns1998efficient}, efficient algorithms for SQ learning of complex models such as DNNs do not exist.}

A naive way to have such models provably forget is to retrain them from scratch. To avoid the large computational and time overhead associated with fully retraining models affected by training data erasure, our research seeks to hold ML to standards such as the right to be forgotten instead through the ability to \textit{unlearn}. Given a trained model, unlearning assures the user that the model is no longer trained using the data which the user elected to erase. \new{Put another way, unlearning guarantees that training on a point and unlearning it afterwards will produce the same distribution of models that not training on the point at all, in the first place, would have produced. }%

Due to this strong definition, we do not consider the setting in which unlearning is used to mitigate poisoning attacks~\cite{nelson_exploiting_nodate,rubinstein_antidote:_2009,biggio2012poisoning}; the guarantee we provide is far stricter than what would be needed for poisoning---\ie that the loss of model accuracy due to the poisoning are mitigated. \new{Instead, we focus on mechanisms that provide the stronger \textit{privacy-minded} unlearning guarantee described above %
in order to satisfy the right to be forgotten requirement. }

Our \SISA approach, short for {\bf S}harded, {\bf I}solated, {\bf S}liced, and {\bf A}ggregated training, can be implemented with minimal modification to existing pipelines. \new{ First, we divide the training data into multiple disjoint shards such
that a training point is included in one shard only;
shards partition the data.
Then, we train models in isolation on each of these shards, which
limits the influence of a point to the model
that was trained on the shard containing the point. 
Finally, when a request to unlearn a training point arrives, we need to retrain only the affected model.  }
Since shards are smaller than the entire
training set, this decreases the retraining time to achieve unlearning.
\new{However, by doing so, we are reducing the amount of data per shard, which may result in a {\em weak learner}~\cite{kearns1988thoughts}.}
In addition, rather than training each model on the entire shard directly, 
we can divide each shard's data into slices and present slices
incrementally during training. We save the state of model parameters
before introducing each new slice, allowing us to start retraining the model
from the last known parameter state  that does not include the point to be unlearned---rather than a random initialization. Slicing 
further contributes to decreasing the time to unlearn, at
the expense of additional storage. \new{At inference, we use different  strategies to aggregate the predictions of models trained on each shard: the simplest one is a majority vote over predicted labels. }

 \new{To demonstrate that \SISA  
handles streams of unlearning requests effectively, we analytically compute speed-ups achieved when the service
provider processes unlearning requests sequentially
(\ie immediately upon a user revoking access to their data)
or in batches (\ie the service provider buffers a few unlearning
requests before processing them). Our results show that \SISA achieves more advantageous trade-offs between accuracy and time to unlearn---compared to two baselines: (1) the naive approach of retraining from scratch, and (2) only train on a fraction of the original training set (\ie only use one of the shards to predict). }

\new{We first turn to {\em simple} learning tasks, such as deep networks trained on Purchase and SVHN. When processing $8$ unlearning requests on Purchase and $18$ unlearning requests on SVHN, we find that \SISA achieves a speed-up of $4.63\times$ and $2.45\times$ over the first baseline---through the combined effect of partitioning the data in $20$ shards each further divided in $50$ slices. This comes at a nominal degradation in accuracy of less than $2$ percentage points. The second baseline is only viable when training on a large fraction $\frac{1}{S}$ of the data: it outperforms \SISA by a factor of $S$ but quickly induces a large cost in accuracy as $S$ increases. Compared to these baselines, we conclude that \SISA enables the service provider to find a more advantageous compromise between model accuracy and time to unlearn.
Next, we turn to more {\em complex} learning tasks involving datasets such as Imagenet and deeper networks. With the same setup (\ie number of shards and slices), we observe a speed-up of 1.36$\times$, at the expense of a greater accuracy degradation ($19.45$ percentage points for top-5 accuracy) for 39 requests\footnote{For 4 requests, observe an 8.01$\times$ speed-up for mini-Imagenet at the expense of 16.7 percentage points accuracy degradation.}. We demonstrate that transfer learning can significantly reduce this accuracy degradation.}

\new{We observe that speed-up gains from sharding exist when the number of unlearning requests is less than three times the number of shards.  However, for complex learning tasks, increasing the number of shards results in a decrease in aggregate accuracy. Slicing, however, always provides a speed-up. While the number of unlearning requests may seem small, these are {\em three orders of magnitude larger} than those in prior work~\cite{bertram2019five}. These savings in retraining times enable large organizations to benefit from economies of scale.}

When faced with different distributions of unlearning requests, \ie requests are not uniformly issued across the dataset, we present a refined variant of our approach, which assumes prior knowledge of the distribution of unlearning requests. We validate it in a scenario that models a company operating across multiple jurisdictions with varying legislation and sensitivities to privacy, and accordingly varying distributions of unlearning requests from users based on publicly available information~\cite{bertram2019five}. 
Knowing this distribution enables us to further decrease expected unlearning time by placing the training points that will likely need to be unlearned in a way that reduces retraining time. \new{For simple learning tasks,} the cost in terms of accuracy is either null or negligible, depending on the distribution of requests considered. 

In summary, the contributions of this paper are:
\begin{itemize}
    \item We formulate a new, intuitive definition of unlearning.  Our definition also takes into account non-uniform distributions of unlearning requests. 
    \item We introduce \SISA, a practical approach for unlearning that relies on data sharding and slicing to reduce the computational overhead of unlearning.
    \item We analytically derive the asymptotic reduction in time to unlearn points with sharding and slicing when the service provider processes requests sequentially or in batches.
    \item We demonstrate that sharding and slicing combined do not impact accuracy \new{significantly for simple learning tasks}, and that \SISA could be immediately applied to handle orders of magnitude more unlearning requests than what Google anticipates is required to implement the GDPR right to be forgotten~\cite{bertram2019five}.
    \item \new{For complex learning tasks, we demonstrate that a combination of transfer learning and \SISA induces a nominal decrease in accuracy ($\sim2$ percentage points) with improved retraining time.}
\end{itemize}

\section{Background on Machine learning}
\label{sec:notation}

We provide rudiments of machine learning as
they apply to neural networks. We chose to study  
neural networks because they almost always generate the largest computational 
costs and require investments in dedicated accelerators~\cite{krizhevsky2012imagenet,jouppi2017datacenter}.

Our efforts fall under the realm of supervised machine learning~\cite{shalev2014understanding}. 
Tasks to be learned are
defined in a space $\mathcal{Z}$ of the form $\mathcal{X} \times \mathcal{Y}$, where $\mathcal{X}$ is the sample space and $\mathcal{Y}$ is the output space.
For example, $\mathcal{X}$ could be thought of as the space of images and $\mathcal{Y}$ as the labels of the images. %

Given a dataset of input-output pairs $(x, y)\in \mathcal{X}\times\mathcal{Y}$, the goal of a supervised learning algorithm is to find
a model, \ie a function $F:\mathcal{X} \mapsto \mathcal{Y}$ that maps these inputs to outputs. 
The learning algorithm that produces this model uses an optimizer. It takes in a dataset, a hypothesis space, and an objective:
\begin{itemize}
\item \textit{Dataset:} Consistent with the probably approximately correct (PAC) learning setting~\cite{valiant1984theory}, we assume there is an underlying distribution on $\mathcal{Z}$ that describes the data; the learner has no direct knowledge of the distribution but has access to a dataset $\mathcal{D}$ that is drawn from it. This dataset $\mathcal{D}$ is further split into the training dataset $\mathcal{D}_{tr}$ and a holdout dataset called the test dataset $\mathcal{D}_{te}$ such that $\mathcal{D}_{te} \cup \mathcal{D}_{tr} = \mathcal{D}$ and $\mathcal{D}_{te} \cap \mathcal{D}_{tr} = \varnothing$. 
\item \textit{Hypothesis space:} An hypothesis is a set of parameter values $w$, which together with the model architecture $F$ selected, represent one possible mapping $F_w:\mathcal{X} \mapsto \mathcal{Y}$ between inputs and outputs. In our case, the hypothesis is a neural network and its parameters are the weights that connect its different neurons (see below).
\item \textit{Objective:} Also known as the loss function, the objective characterizes how good any hypothesis is by measuring its empirical risk on the dataset, \ie approximate the error of the model on the underlying task distribution of which we only have a few samples.
A common example is the cross-entropy loss, which measures how far a model's outputs are from the label: $l(x,y)= -\sum_{i=0}^{n-1} y_i \cdot \log(F_w(x))$ where $n$ is the number of classes in the problem.
\end{itemize}
Given an architecture $F$, a model $F_w$ is found by searching for a set of parameters $w$ that 
minimize the empirical risk of $F_w$ on the training set $\mathcal{D}_{tr}$.  Performance of the model is validated by measuring its accuracy on the test dataset $D_{te}$.

We experiment with our approach using neural networks and deep learning~\cite{lecun2015deep}. Deep neural networks (DNNs) are non-parametric functions organized as layers. Each layer is made of {\em neurons}---elementary computing units that apply a non-linear activation function to the weighted average of their inputs. Neurons from a given layer are connected with weights to neurons of the previous layer. The layout of these layers and the weight vectors that connect them constitutes the {\em architecture} of the DNN, while the value of each individual weight (collectively denoted by $w$) is to be learned. Weights are updated using the {\em backpropagation} algorithm~\cite{rumelhart1986learning}. The algorithm starts by assigning a random value to each weight. Then, a data point is sampled from the dataset and the loss function is computed to compare the model's prediction to the data point's label. Each model parameter value is updated by multiplying the gradient of the loss function with respect to the parameter by a small constant called the learning rate. This algorithm enables learning and gradually improves the model's predictions as more inputs are processed.

\section{Defining Unlearning}
\label{sec:definition}

A requirement of privacy regulations such as the GDPR or the CCPA is that individuals whose data is housed by organizations have the right to request for this data to be erased.  This requirement poses challenges to current machine learning technologies. We define the \textit{unlearning problem} by examining these challenges, which then leads us to a formal model of the unlearning problem.  We identify objectives for an effective approach to unlearning, which we use to show the ineffectiveness of existing {\em strawman} solutions.

\subsection{Why is Unlearning Challenging?}

The reason unlearning is challenging stems from the complex and stochastic nature of training methods used \new{to optimize model parameters  in modern ML pipelines. }

\vspace{1mm}
\noindent{\em 1. We have a limited understanding of how each data point impacts the model}. There exists no prior work that measures the influence of a {\em particular training point on the parameters of a model}. While research has attempted to trace a particular test-time prediction through the model's architecture and back to its training data~\cite{cook1980characterizations,koh2017understanding}, these techniques rely on influence functions, which involve expensive computations of second-order derivatives of the model's training algorithm. Further, it is not obvious how to modify such influence functions so that they map the effect of a single training point on model parameters \new{for complex models such as DNNs}. \new{We later discuss techniques for differentially private learning, which seek to bound the influence any training point can have on model parameters, and explain how they are inadequate because the bound is always non-zero. }

\vspace{1mm}
\noindent{\em 2. Stochasticity in training.} A great deal of randomness exists in the training methods for complicated models such as DNNs; small batches of data (\eg with 32 points) are randomly sampled from the dataset, and the ordering of batches varies between different epochs, \ie passes of the algorithm through the dataset.  Further, training is often parallelized without explicit synchronization, meaning the inherent random ordering of parallel threads may make the training non-deterministic.

\vspace{1mm}
\noindent{\em 3. Training is incremental.}
Additionally, training is an incremental procedure where any given update reflects all updates that have occurred prior to it. For example, if a model is updated based on a particular training point (in a particular batch) at a particular epoch, all subsequent model updates will depend, in some implicit way, on that training point.

\vspace{1mm}
    \noindent{\em 4. Stochasticity in learning.} Intuitively, learning algorithms are designed to search for an optimal hypothesis in a vast hypothesis space. In the case of neural networks, this space contains all models that can be defined by setting the weights of a fixed neural network architecture. PAC learning theory suggests that the learned hypothesis is one of many hypotheses that minimize the empirical risk. For example, the common choice of optimizer for neural networks, stochastic gradient descent, is capable of converging to one of the many local minima for any convex loss function. Coupled with the stochasticity involved in training, it is very challenging to correlate a data point with the hypothesis learned from it.

\subsection{Formalizing the Problem of Unlearning}

We formalize the unlearning problem as a game between two entities: an \new{\em honest} service provider $\mathcal{S}$, and a user population $\mathcal{U}$. The service provider could be a large organization that collects information from various individuals (such as a company or hospital). This data is curated in the form of a dataset $\mathcal{D}$. The service provider uses this data for training and testing a machine learning model $M$ in any way they desire. Any user $u \in \mathcal{U}$ can revoke access to their individual data $d_u \subset \mathcal{D}$. Observe that $d_u$ can be a single element in the dataset, or a set of elements. Within a finite period of time, the service provider has to erase the revoker's data and modify any trained models $M$ to produce $M_{\lnot d_u}$, where $M_{\lnot d_u}$ is some model that could plausibly have been trained if $d_u$ were not in $\mathcal{D}$. \new{In Definition~\ref{def:unlearning}, we define plausibility according to the distribution of models output by the training algorithm.}  Further, $\mathcal{S}$ must convince $u$ that $M_{\lnot d_u}$ is such a model---a defense akin to that of \textit{plausible deniability}.  Access to data may be revoked by users sequentially, but the service provider may choose to perform data erasing in a batched fashion, as discussed in \S~\ref{sec:evaluation}.

\begin{figure}[t]
\centering
\includegraphics[width=\linewidth]{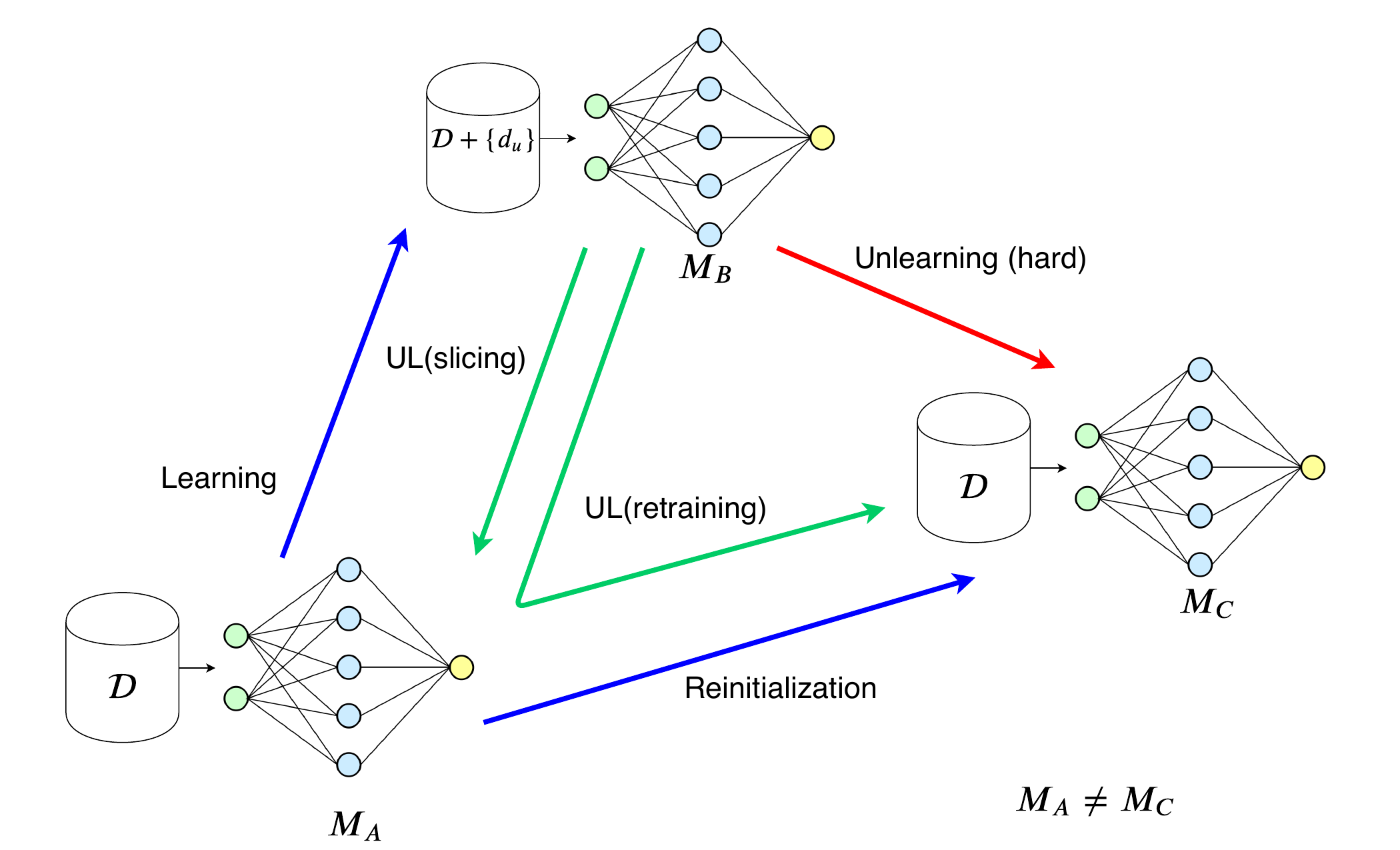}
\caption{\small Unlearning (\underline{\textcolor{red}{red arrow}}) is hard because there exists no function that measures the influence of augmenting the dataset $\mathcal{D}$ with point $d_u$ and fine-tuning a model $M_A$ already trained on $\mathcal{D}$ to train (\underline{\textcolor{blue}{left blue arrow}}) a model $M_B$ for $\mathcal{D}+\{d_u\}$. This  makes it impossible to revert to model $M_A$ without saving its parameter state before learning about $d_u$. We call this model slicing (\underline{\textcolor{green}{short green arrow}}). In the absence of slicing, one must retrain (\underline{\textcolor{green}{curved green arrow}}) the model without $d_u$, resulting in a model $M_C$ that is different from the original model $M_A$.\vspace{-5mm}}
\label{fig:unlearning}
\end{figure}

We illustrate this scenario in Figure~\ref{fig:unlearning}. \new{One can observe that given a dataset $\mathcal{D}$, it is possible to train one of several models (\eg DNNs) that generalize well from this dataset unless the learning hypothesis class leads to a unique closed form solution (\eg linear classifier).} We denote two such models $M_A$ and $M_C$. If we add one more data point $d_u$ to the dataset $\mathcal{D}$, we can train another model on this new dataset $\mathcal{D'}$ in many ways. This includes using the parameters of $M_A$ to initialize a new model (rather than randomly initializing it) and continuing training from there on to obtain model $M_B$. Since there is no \new{efficient} function that measures the influence of this one additional point $d_u$ on the parameters in $M_B$, it is very hard to invert the procedure unless a copy of $M_A$ had been previously saved. Later in \S~\ref{sec:approach}, we will define this strategy, termed \textit{slicing}. In the absence of slicing, the most convincing way to obtain plausible deniability, and ensure that the model is devoid of the influence of a particular training point $d_u$, is to retrain it from scratch without that particular point (keeping all other training hyperparameters the same) \ie use $\mathcal{D'} \setminus d_u$ to obtain the model $M_C$ in our example from Figure~\ref{fig:unlearning}.  It is conceivable that the parameters of $M_A$ and $M_C$ are similar (despite stochasticity in learning) and it is desired for their performance (in terms of test accuracy) to be comparable. However, the fact that  model $M_C$ was obtained by training on $\mathcal{D'} \setminus d_u$ {\em from scratch} provides a certificate to the data owner that their data share was indeed removed. This conveys a very strong notion of privacy.

\begin{definition}
\new{Let $\mathcal{D}=\{d_i: i\in \mathcal{U}\}$ denote the training set collected from population $\mathcal{U}$. Let $\mathcal{D'} = \mathcal{D} \cup d_u$. Let $\mathbb{D}_{\mathcal{M}}$ denote the distribution of models learned using mechanism $\mathcal{M}$ on $\mathcal{D'}$ and then unlearning $d_u$. Let $\mathbb{D}_{real}$ be the distribution of models learned using $\mathcal{M}$ on $\mathcal{D}$. The mechanism $\mathcal{M}$ facilitates unlearning when these two distributions are identical.}
\label{def:unlearning}
\end{definition}

We draw the attention of the reader to two key aspects of the definition. First, the definition captures inherent stochasticity in learning: it is possible for multiple hypotheses to minimize empirical risk over a  training set. As illustrated by models $M_A$ and $M_C$ in Figure~\ref{fig:unlearning}, two models having different parameters does not imply that they were trained with a different dataset. Conversely, two models trained with a different dataset do not necessarily have different parameters.
Second, the definition does not necessarily require that the owner retrain the model $M'$ from scratch on $\mathcal{D}\setminus d_u$, as long as they are able to provide evidence that model $M'$ could have been trained from scratch on $\mathcal{D}'\setminus d_u$. \new{In our work, this evidence takes the form of a training algorithm, which if implemented correctly, guarantees that the distributions $\mathbb{D}_{\mathcal{M}}$ and $\mathbb{D}_{real}$ are identical.}

\subsection{Goals of Unlearning}
\label{ssec:owner-goals}

The simple strategy we have discussed thus far \ie training a model from scratch on the dataset without the point being unlearned is very powerful. We refer to this strategy as the {\em baseline} strategy through the rest of the paper. However, for large dataset sizes, such an approach will quickly become intractable (in terms of time and computational resources expended). For example, to be compliant with GDPR/CCPA, organizations will have to retrain models very frequently. Thus, any new strategy should meet the following requirements.

\begin{enumerate}
    \item[G1.] {\em Intelligibility:} Conceptually, the baseline strategy is very easy to understand and implement. Similarly, any unlearning strategy should be intelligible; this requirement ensures that the strategy is easy to debug by non-experts. 
    \item[G2.] {\em Comparable Accuracy:} It is conceivable that the accuracy of the model degrades, even in the baseline, if (a) the fraction of training points that need to be unlearned becomes too large, or (b) \new{prototypical points~\cite{kim2014bayesian} are unlearned}. Even if there is no component of the approach that explicitly promotes high accuracy, any unlearning strategy should \new{strive to} introduce a small accuracy gap in comparison to the baseline for any number of points unlearned.
    \item[G3.] {\em Reduced Unlearning Time:} The strategy should have provably lower time than the baseline for unlearning any number of points. 
    \item[G4.] {\em Provable Guarantees:} Like the baseline, any new strategy should provide provable guarantees that any number of points have been unlearned (and do not influence model parameters). Additionally, such a guarantee should be intuitive and easy to understand for non-experts~\cite{saltzer1975protection}.
    \item[G5.] {\em Model Agnostic:} The new strategy for unlearning should be general \ie should provide the aforementioned guarantees for models of varying nature and complexity.
    \item[G6.] {\em Limited Overhead:} Any new unlearning strategy should not introduce additional overhead to what are already computationally-intense training procedures. 
\end{enumerate}

\subsection{Strawman Solutions}
\label{strawman}

Based on the requirements discussed earlier, we propose several strawman candidates for an unlearning strategy. The goals specified (sometimes in parantheses) are the goals the strawman solutions {\em do not} meet.

\vspace{1mm}
\noindent{\em 1. Differential Privacy:} Proposed by Dwork \etal~\cite{dwork2011differential}, $\varepsilon$-differential privacy offers probabilistic guarantees about the privacy of individual records in a database. \new{In our case, $\varepsilon$ bounds the changes in model parameters that may be induced by any single training point. While several efforts~\cite{abadi2016deep,chaudhuri2009privacy} make it possible to learn with differential privacy, 
this guarantee is different from what we wish to provide. We require that a point has no influence on the model once it has been unlearned. While differential privacy allows us to bound the influence any point may have on the model, that bound remains non-zero. This implies that there is a possibility that a point still has a small but non-zero influence on the model parameters. To guarantee unlearning, we would need to achieve $\varepsilon$-differential privacy with $\varepsilon=0$. This would make it impossible for the algorithm to learn from the training data ({\bf G2}).}

\vspace{1mm}
\noindent{\em 2. Certified Removal Mechanisms:} Other mechanisms relax the definition of differential privacy to provide certificates of data removal. This includes two concurrent proposals~\cite{guo2019certified,golatkar2020eternal} \new{The mechanism by Guo \etal~\cite{guo2019certified} uses a one-step Newton update~\cite{koh2017understanding}. While such a mechanism introduces a small residue, this is masked by adding noise (similar to approaches in differential privacy). However, as before, their guarantees are probabilistic, and different from what we wish to provide with \SISA. Additionally, to train non-linear models, they resort to pretraining models on public data (for which no guarantees are provided) or from differentially-private feature extractors. In summary, such a mechanism is effective for simple models such as linear regression models, which suggest that they fall short of achieving \textbf{G5}.}

\vspace{1mm}
\noindent{\em 3. Statistical Query Learning:} Cao \etal~\cite{cao_towards_2015} model unlearning in the statistical query learning framework~\cite{kearns1998efficient}. By doing so, they are able to unlearn a point when the learning algorithm queries data in an order decided prior to the start of learning. In this setting, it is possible to know exactly how individual training points contributed to model parameter updates. However, their approach is not general\footnote{\new{Kearns~\cite{kearns1998efficient} shows that any PAC learning algorithm has a corresponding SQ learning equivalent. However, an efficient implementations of SQ equivalents for more complex algorithms does not exist, to the best of our knowledge.}} ({\bf G5}) and does not easily scale to more complex models (such as those considered in this work). Indeed, these models are trained using adaptive statistical query algorithms which make queries that depend on all queries previously made. In this setting, the approach of Cao \etal~\cite{cao_towards_2015} diverges in an unbounded way unless the number of queries made is small, which is not the case for the deep neural networks we experiment with. 

\vspace{1mm}
\noindent{\em 4. Decremental Learning:}  Ginart \etal~\cite{ginart_making_2019} consider the problem from a data-protection regulation standpoint. They present a formal definition of complete data erasure which can be relaxed into a distance-bounded definition. Deletion time complexity bounds are provided. They note that the deletion and privacy problems are orthogonal, which means deletion capability does not imply privacy nor vice versa. However, it is unclear if the approach presented (Quantized k-Means) is applicable ({\bf G5}) and scalable ({\bf G6}) for all model classes.

\section{The \SISA Approach}
\label{sec:approach}

Our discussion thus far motivates why retraining from scratch while omitting data points that need to be unlearned is the most straightforward way to provide provable guarantees. However, this naive strategy is inefficient in the presence of large datasets or models with high capacity that take a long time to train. We present, {\bf SISA} (or {\bf S}harded, {\bf I}solated, {\bf S}liced, {\bf A}ggregated) training to overcome these issues.

\begin{figure}[t]
\centering
\includegraphics[width=\linewidth]{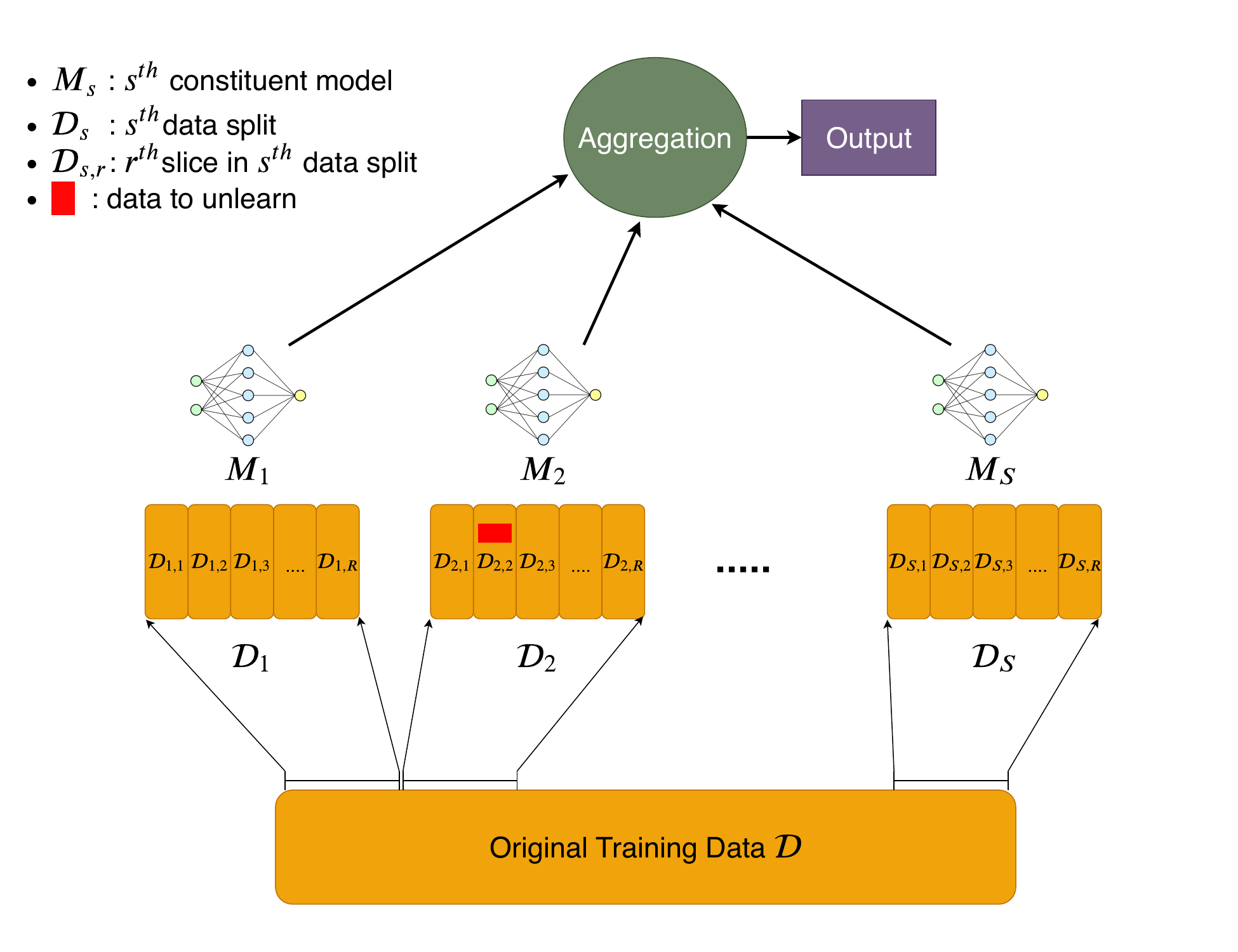}
\caption{\small \SISA: data is divided in shards, which are themselves divided into slices. One constituent model is trained on each shard by presenting it with incrementally many slices and saving its parameters before the training set is augmented with a new slice. When data needs to be unlearned, only one of the constituent models whose shards contains the point to be unlearned needs to be retrained --- retraining can start from the last parameter values saved before including the slice containing the data point to be unlearned.\vspace{-4mm}}
\label{fig:sisa}
\end{figure}

\subsection{The \SISA Approach to Training}

As illustrated in Figure~\ref{fig:sisa}, \SISA replicates the model being learned several times where each replica receives a {\em disjoint shard} (or subset) of the dataset---similar to current distributed training strategies~\cite{dean2012large,ben2019demystifying}. We refer to each replica as a constituent model. However, \SISA deviates from current strategies in the way incremental model updates are propagated or shared—{\em there is no flow of information between constituent models}.  For example, if each constituent model is a DNN trained with stochastic gradient descent, then gradients computed on each constituent are not shared between different constituents; each constituent is trained in isolation. This ensures that the influence of a shard (and the data points that form it) is restricted to the model that is being trained using it. Each shard is further partitioned into slices, where each constituent model is trained incrementally (and iteratively, \new{in a stateful manner}) with an increasing number of slices. At inference, the test point is fed to each constituent and all the constituents' responses are aggregated, similar to the case of ML ensembles~\cite{dietterich2000ensemble}.
 
 When a data point is to be unlearned, only the constituent model whose dataset contains this point is affected. More specifically, a data point is unlearned from a particular slice in a particular shard. Retraining can start from the last parameter state saved prior to including the slice containing the data point to be unlearned: only the models that are trained using the slice containing the unlearned point need to be retrained. We will describe each component in more detail in \S~\ref{techniques}. 

Observe that \new{our analysis of unlearning however assumes that the retraining time grows linearly in the size of the dataset. We validate this assumption in \S~\ref{linear time validation}}. However, we make no assumptions about the nature of the constituent models or if the constituents are homogeneous (\ie the same model or hypothesis class) or heterogeneous (\ie different models or hypothesis class). Sharding is possible for any
model or hypothesis class: it has no impact on how training is performed beyond 
the smaller set of data each model has access to. Slicing is possible for
any iterative learning algorithm that is stateful: the algorithm should be
such that it can continue to learn from its current state when presented with new
data. Gradient descent naturally falls under that category. 
However, decision tree learning is a counter-example of a technique
that does not benefit from slicing, because 
it greedily chooses a feature to add to the decision tree based
on how well it splits the data according to a metric like Gini impurity. 
For this reason, when a new slice of data is added, the tree
must be constructed again from scratch. In summary, slicing can be 
used for any model that is trained through gradient descent: \eg logistic regression and neural networks, but also support vector machines in some cases~\cite{shalev2011pegasos}.

The key requirement of our training strategy is that the the updates obtained during the iterative training process are not exchanged between different constituents. Intuitively, such an approach may seem detrimental to improving the generalization capabilities of the model; each constituent is trained on a (significantly) smaller portion of the dataset, \new{and may become a \textit{weak learner}~\cite{kearns1988thoughts}. We evaluate this aspect in \S~\ref{sec:evaluation}, and discuss trade-offs of several aggregation approaches to mitigate this effect for different learning tasks.}

\subsection{Techniques}
\label{techniques}

\noindent{\em 1. Sharding:} By dividing the data into disjoint fragments and training a constituent model on each smaller data fragment, we are able to  distribute the training cost. While this means our approach naturally benefits
from parallelism across shards, we do not take this into account in our
analysis and experiments, out of fairness to the baseline of retraining a 
model from scratch (which could also be accelerated by distributing
the computation across multiple machines). 

For the remainder of this section, we assume that we have no prior information associated with the probabilities with which each individual point might be unlearned. In such a scenario, a dataset $\mathcal{D}$ can be uniformly partitioned into $S$ shards such that $\cap_{k \in [S]} \mathcal{D}_k = \varnothing$ and $\cup_{k \in [S]} \mathcal{D}_k = \mathcal{D}$. For each shard $\mathcal{D}_k$, a model (denoted $M_k$) is trained using the entirety of the data available in $\mathcal{D}_k$. In \S~\ref{sec:ulrequests}, we explore the scenario where the distribution
of unlearning requests is known to $\mathcal{S}$. 

Observe that user $u$'s data-point $d_u$ can lie in each of the $S$ shards with equal probability.  Moreover, one of the parameters of the training can be whether each $d_u$ be part of only one shard or several.  For simplicity, we will assume that each $d_u$ is part of only one shard, as this maximizes the savings in unlearning time.  We discuss this further in \S~\ref{sec:discussion}. If the user desires for $d_u$ to be unlearned, then the service provider has to (a) first locate the dataset (and shard) in which $d_u$ is located, referred to as $\mathcal{D}_u$, and (b) retrain {\em from scratch} the corresponding model on $\mathcal{D}_u$ \textbackslash $d_u$; this will result in a new model $M'_u$. In comparison, the baseline would entail retraining the model {\em from scratch} on $\mathcal{D}$ \textbackslash $d_u$. Since $|\mathcal{D}| >> |\mathcal{D}_u|$, the time required for retraining (henceforth referred to as retraining time) in the baseline is far greater than in our proposal; our proposal provides an expected speed-up of $S\times$.\footnote{For a single unlearning request.}

\vspace{1mm}
\noindent{\em 2. Isolation:} Observe that based on the proposal detailed earlier, the training of each shard occurs in isolation. %
\new{By not performing a joint update,} we potentially degrade the generalization ability of the overall model (comprising of all constituents). However, we demonstrate that for appropriate choices of the number of shards, this does not occur in practice \new{for certain types of learning tasks}. Isolation is a subtle, yet powerful construction that enables us to give concrete, provable, and intuitive guarantees with respect to unlearning.

\vspace{1mm}
\noindent{\em 3. Slicing:} By further dividing data dedicated for each model (\ie each shard) and incrementally tuning (and storing) the parameter state of a model, we  obtain additional time savings. Specifically, each shard's data $\mathcal{D}_k$ is further  uniformly partitioned into $R$ disjoint {\em slices} such that $\cap_{i \in [R]} \mathcal{D}_{k,i} = \varnothing$ and $\cup_{i \in [R]} \mathcal{D}_{k,i} = \mathcal{D}_k$. We perform training for $e$ epochs to obtain $M_k$ as follows: 
\begin{enumerate}
    \item At step 1, train the model {\em using random initialization} using only $\mathcal{D}_{k,1}$, for $e_1$ epochs. Let us refer to the resulting model as $M_{k,1}$. Save the state of parameters associated with this model.
    \item At step 2, train the model $M_{k,1}$ using $\mathcal{D}_{k,1} \cup \mathcal{D}_{k,2}$, for $e_2$ epochs. Let us refer to the resulting model as $M_{k,2}$. Save the parameter state.
    \item At step $R$, train the model $M_{k,R-1}$ using $\cup_i \mathcal{D}_{k,i}$, for $e_R$ epochs. Let us refer to the resulting {\em final} model as $M_{k,R} = M_k$. Save the parameter state.
\end{enumerate}

As before, observe that if user $u$'s data-point $d_u$ lies in shard $\mathcal{D}_k$, then it can lie in any of the $R$ slices with equal probability. Thus, if the user desires for $d_u$ to be unlearned, then the service provider has to (a) first locate the slice in which $d_u$ is located, referred to as $\mathcal{D}_{k,u}$, and (b) perform the training procedure as specified above from step $u$ onwards using $\mathcal{D}_{k,u}$ \textbackslash $d_u$; this will result in a new model $M'_{k,u}$. \new{For a single unlearning request,} \new{this provides a best-case speed-up up to $\frac{R+1}{2}\times$ compared to using the strategy} without slicing (\new{we discus this in more detail in \S~\ref{time_slice}}).

It is also worth noting that the duration of training for the constituent models with and without data slicing is different when they have the same number of epochs. Each epoch takes less time when only a subset of the slices is being trained on; on the other hand, training incremental combinations of slices takes longer because the training process is repeated after each slice is added. In order to train models with and without slicing {\em for the same amount of time}, we introduce the following relationship between the number of epochs with and without slicing. Let $D = \frac{N}{S}$ be the number of points per shard, where $N$ is the size of the dataset. Let $e'$ be the number of epochs without slicing; we seek to find the number of epochs $e=\sum_{i=1}^R e_i$ to train a model with $R$ slices, where $e_i$ is the number of epochs required to train $\frac{iD}{R}$ samples. We make a simplifying assumption: we assume that each slice is trained equally long \ie $\forall i, e_i = \frac{e}{R}$. We also assume that the training time is estimated solely based on the amount of training data (as detailed in \S~\ref{sec:time}).

\begin{equation}
\label{eq:slice_epoch}
\new{e'D =  \sum_{i=1}^{R} e_i\frac{iD}{R} \equiv e = \frac{2R}{R+1}e'}
\end{equation}

The speed-up provided by slicing comes at no expense beyond the overhead induced by storing the state of parameters before each slice is introduced in training. \new{We explore these trade-offs in detail in Appendix~\ref{app:storage}.}

\vspace{1mm}
{\em 4. Aggregation:} At inference time, predictions from various constituent models can be used to provide an overall prediction. \new{The choice of aggregation strategy in \SISA is influenced by two key factors:} 
\begin{enumerate}
\item \new{It is intimately linked to how data is partitioned to form shards: the goal of aggregation is to maximize the joint predictive performance of constituent models.}%
\item \new{The aggregation strategy should not involve the training data (otherwise the aggregation mechanism itself would have to be unlearned in some cases).} 
\end{enumerate}
\new{In the absence of knowledge of which points will be the subject of unlearning requests, there is no better strategy than to partition\footnote{\new{Partition applies to both shards and slices here.}} data uniformly and to opt for a voting strategy where each constituent contributes equally to the final outcome through a simple {\em label-based majority vote}. This naturally satisfies both requirements above.} %

\new{In cases where constituent models assign high scores to multiple classes rather than a single class, the majority vote aggregation loses information about the runner-up classes. In \S~\ref{big}, we evaluate a refinement of this strategy where we average the entire prediction vectors (\ie the post-softmax vector indicating the model's confidence in predicting each class) and pick the label of the highest value. 
We also considered training a controller model that re-weights predictions made by constituent models~\cite{shazeer2017outrageously}, \ie that learns which model is best for predicting on a given test point. However improvements in accuracy were modest and not worth the cost of now having to retrain the controller model if its own training data is the subject of an unlearning request later made by a user. }

\vspace{1mm}
\textbf{Take-away.} In summary, the techniques discussed here can provide \new{a best-case speed-up of $\frac{(R+1)S}{2}\times$ in} terms of retraining time (for one unlearning request). %
However, our approach introduces several challenges.

\subsection{Challenges}

We make no assumptions about (a) the nature of unlearning requests, (b) the nature of training algorithms, and (c) the nature of data distribution within both the shards and slices. This results in several challenges which we discuss below. 

\subsubsection{Weak Learners} \new{We motivate the notion of weak learners with the concept of {\em task complexity}\footnote{\new{The notion of task complexity is subjective. For example, if MNIST is considered a simple task, few shot learning~\cite{snell2017prototypical} of MNIST can be complex.}} -- defined as a function of (a) the input dimensionality, (b) the complexity of the model (in our case, DNN) used to solve a particular learning task, and (c) the number of samples per class available to the model for learning. Datasets such as MNIST~\cite{mnist} are considered to be simple because they (a) have inputs with few features, (b) are trained over deep neural networks with few hidden layers, and (c) have a large number of samples per class. Instead, Imagenet~\cite{imagenet} is considered complex with over 150,000 features and 1000 classes: it requires neural networks with a large number of hidden layers (in the order of a 100s).}

Since each constituent model is trained on a small shard, these models could be weak learners~\cite{kearns1988thoughts,freund1997decision}: in other words, their accuracy will be lower than a single model trained on the entire dataset. \new{This effect is more profound in {\em complex} learning tasks.} The primary reason for why this accuracy gap could exist is that when each constituent model is trained on very limited data \new{which is also not prototypical~\cite{kim2014bayesian}---especially when the number of samples per class is low;} if the model has high-capacity (as is the case with DNNs), the model might overfit to the small training dataset. 

Some of this accuracy will be recovered by the aggregation operation. However, we instantiate our approach assuming that the constituent models trained on shards are all trained with the same architecture, and the same hyperparameters. Since this departs 
from prior work on ML ensembles, which typically involves an ensemble of heterogeneous models~\cite{opitz1999popular} trained with different techniques, we may not obtain as large benefits from aggregation as  is typically the case.

\subsubsection{Hyperparameter Search}

Additionally, sharding and slicing may require that the service provider
revisit some hyperparameter choices made on the entire dataset. For instance,
sharding and slicing may require training with a different number of epochs. 
Slicing could also negatively interact with batching when the service
provider is using a large number of slices---because each slice will be smaller. 

If each constituent model requires a different set of hyperparameters for optimal performance, then as the number of models (of the order $O(SR)$) increases, performing hyperparameter tuning is a truly challenging problem.
Training $O(SR)$ models, depending on the hyperparameter search needed to optimize for these challenges, may introduce a computational overhead. 
We note that hyperparameters are shared across constituent models 
when data is split uniformaly across shards. In that case, one only needs
to train $O(R)$ models to tune the hyperparameters for slicing.

\vspace{1mm}
\textbf{Take-away.} We revisit these challenges in \S~\ref{sec:evaluation}, discuss the various solutions we explored for each of the problems listed above, and highlight insights we gained from them.

\section{Measuring Time}
\label{sec:time}

\subsection{Measuring time analytically}
\label{linear time validation}
\begin{figure}
\centering
\subfloat[{{\small SVHN dataset}}]{\label{fig:svhnlinear}{\includegraphics[width=0.5\linewidth]{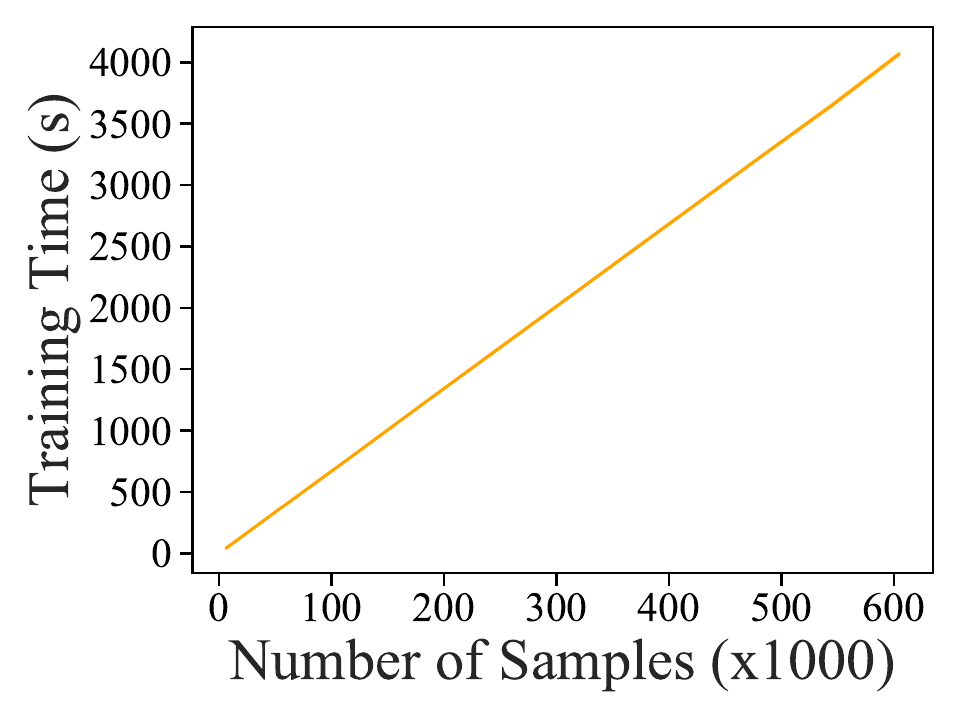}}  } 
\subfloat[{{\small Purchase dataset}}]{\label{fig:purchaselinear}{\includegraphics[width=0.5\linewidth]{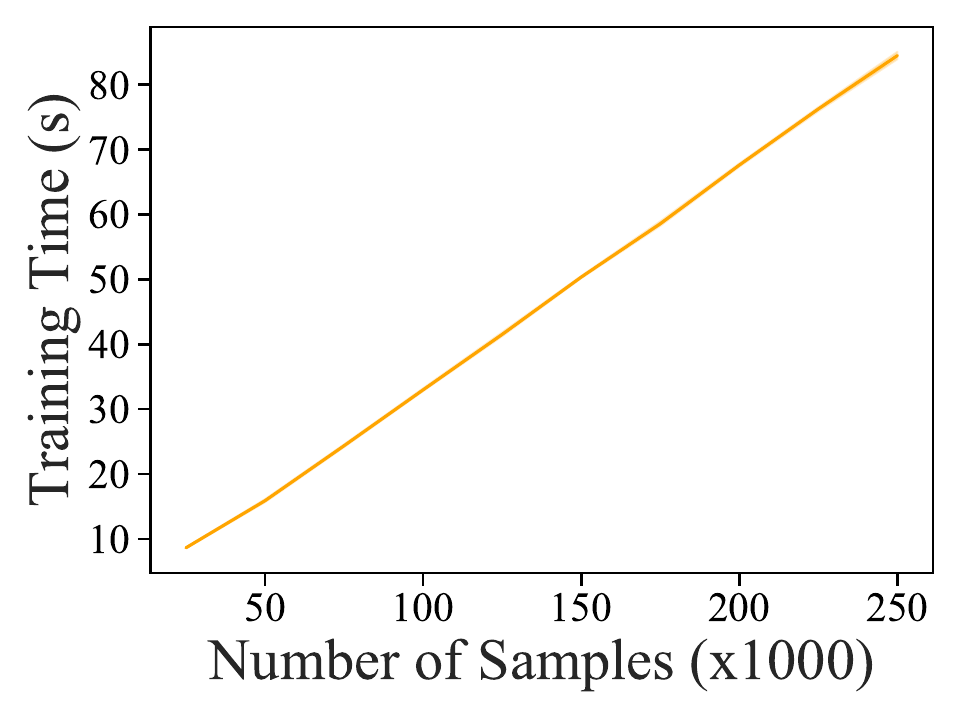}}  } 

\caption{\small \new{We validate the linear relationship (within error) between training time and the number of samples trained on. Measurements are obtained on increments of 10\% of the dataset size.  We repeat 5 times to report mean and variance, on SVHN and Purchase.}\vspace{-3mm}} 
\label{fig:linear_relation}
\end{figure}

\noindent{\bf Motivation.} \new{Measuring time experimentally is difficult because both hardware and software introduce variance in measurements. To circumvent these variances, we measure unlearning time indirectly through the number of samples that one needs to retrain. We were able to validate, in a controlled experiment, the linear relationship between the number of (re)training samples and a model’s training time. This experiment was performed on a workstation equipped with a RTX2080 Ti accelerator and repeated 5 times to estimate variance. For the SVHN and Purchase datasets (described in \S~\ref{datasets}), the results in Figure~\ref{fig:linear_relation} show that the number of samples to retrain is proportional to the retraining time. Note that we verify this relationship for the MNIST dataset as well, but omit the figure due to space constraints.}

Having established this relationship, the following analysis calculates the expectation of the number of data points needed for retraining, given an unlearning request, as the number of shards and slices vary. 

\subsection{Measuring Time for Sharding}
\label{time_shard}

Observe that for each unlearning request, a single constituent model is retrained when it arrives sequentially whereas multiple models are retrained when the requests are batched. 

\vspace{1mm}
\noindent{\em 1. Sequential Setting:} In the sequential setting, we make two assumptions: (a) the training data is shuffled and evenly split into $S$ shards, and (b) each unlearning request can require any of the $S$ shards to be retrained, with equal probability, at any step. %
We explicitly calculate the expectation of the number of points needed to be used for retraining. \new{To achieve our desired result, we make a simplifying assumption: the shard sizes stay roughly the same as points are removed due to unlearning.}

If the sharding is uniform, then each model has (roughly) the same number of initial training data points $\frac{N}{S}$; it is obvious that the first unlearning request will result in retraining of $\frac{N}{S}-1$ points for the {\em one} shard that is affected. For the second unlearning request, there will be two cases: the shard affected in the first unlearning request is affected again, which will result in retraining $\frac{N}{S} - 2$ data points with a probability $\frac{1}{S}$, or any other shard is impacted resulting in retraining $\frac{N}{S} - 1 $ data points with probability $1-\frac{1}{S}$. Thus, inductively, we can see that for the $i^{th}$ unlearning request, the probability that $\frac{N}{S} - 1- j$ points (for $0 \leq j \leq i-1$) are retrained is 
{\small\[{i -1  \choose j} \left( \frac{1}{S} \right)^j \left( 1-\frac{1}{S} \right)^{i-j-1}\]}

\new{By first summing over all possible combinations of points that are unlearned in a shard at a specific step, and then summing over all requests ($K$ in total), we are able to obtain the expected number of points to be retrained ($\E(C)$) as:}
\new{{\small\[\sum_{i=1}^{K}\sum_{j=0}^{i-1} {i-1 \choose j} \left(\frac{1}{S}\right)^j\left(1-\frac{1}{S}\right)^{i-j-1} \left(\frac{N}{S} - 1 - j\right)\]}
}

\new{This expression can be simplified using the binomial theorem, as described in Appendix~\ref{appendix:time_sharding_sequential} to obtain:} 
\new{{\small\begin{equation}
    \E[C] = \left(\frac{N}{S}+\frac{1}{2S}-1\right)K - \frac{K^2}{2S}
\end{equation}}
}

\new{ An upper bound for the above equation can be obtained if we assume that after each unlearning request, the size of each shard remains constant. Thus, the cost of any step is $\frac{N}{S}$. We then have a linear bound for the total cost: $\frac{N}{S}K$; doubling the number of shards involves dividing the number of data points that need retraining by two. This bound captures the behavior of the expected cost when two conditions are met: (a) $K \rightarrow 0$, and (b) $\frac{N}{S} >> 1$. Conversely, for $K \rightarrow N$, the quadratic behavior becomes preponderant.}

\vspace{1mm}
\noindent{\em 2. Batch Setting:} \new{Alternatively, service provider $\mathcal{S}$ could aggregate unlearning requests into a batch, and service the batch.} The cost of unlearning the batch is $C= \sum_{j=1}^{S}{(\frac{N}{S}-\mathfrak{u}_j)\mathfrak{h}_j}$ where $(\mathfrak{u}_j)_{j \in \{1,\dots,S\}}$ are the random variables which indicate the number of times a shard of index $j$ is impacted, and $(\mathfrak{h}_j)_{j \in \{1,\dots,S\}}$ are the Bernouilli random variables indicating if a shard of index $j$ is impacted by an unlearning request. We can show that $(\mathfrak{u}_j)_{j \in \{1,\dots,S\}}$ follows a binomial distribution $B(K,\frac{1}{S})$. Thus, the expected cost is:
{\small\begin{equation}
    \E[C] = N\left(1-\left(1-\frac{1}{S}\right)^K\right)-K
\label{eq: sharding_batch_time_analysis}
\end{equation}}

Asymptotically, $\E[C] \sim N(1-\exp({\frac{K}{\tau})})$ where $\tau = (-\ln({1-\frac{1}{S}}))^{-1}$ when $K \rightarrow 0$, and $\E[C] \sim N - K$ when $K \rightarrow +\infty$. Thus, the benefits of sharding are most noticeable when $K \ll N$ \new{(refer to Appendix~\ref{appendix:time_sharding_batch} for more details)}.

\subsection{Measuring Time for Slicing}
\label{time_slice}

Our analysis of slicing differs from the analysis we presented for sharding because unlike shards, which are independent, a slice depends on all slices observed before them. Again, we distinguish two cases: in the first,  the service provider processes unlearning requests sequentially, and in the second, requests are processed in batches. 

\vspace{1mm}
\noindent{\em 1. Sequential Setting:} The case where unlearning requests are processed  as a stream is easier to analyze. Since we assume that the time for retraining a model is proportional to the number of points needed to be retrained, we need to find the expectation of the number of samples that will need to be retrained for a single unlearning request.

 Recall from \S~\ref{sec:approach} that if an unlearning request happens in the $r^{th}$ slice, we need to retrain all the way to the $R^{th}$ slice. From equation~\ref{eq:slice_epoch}, the expected number of samples that need to retrain is:

\begin{equation}
    \label{eq:slice_seq}
    \E[C] = \E\left[\sum_{i=r}^R\frac{2e'}{R+1}\frac{iD}{R}\right] = e'D\left(\frac{2}{3} + \frac{1}{3R}\right)
\end{equation}
which is an upper bound on the expected number of points to be retrained for a single unlearning request. The upper bound is due to the approximation we make about the number of points per slice $\frac{D}{R}$ remaining constant throughout unlearning. In \S~\ref{sec:evaluation}, we show that this approximation is acceptable when $K \ll N$.
\new{
With $R \rightarrow +\infty$, we have $\E[C] \rightarrow \frac{2}{3}e'D$, which gives the maximum expected speed-up of $1.5\times$. With $R =1$, we have $\E[C] = e'D$ (or no speed-up).} %

\vspace{1mm}
\noindent{\em 2. Batch Setting:} As before, we denote the
number of unlearning requests processed in a batch as $K$. In this case, we need to find the expected 
minimum value over multiple draws of a random variable to compute the index of the slice from which 
we will have to restart training. Each unlearning request can still be modelled
as a random draw from a uniform distribution \new{$U(1, D)$}. 
However, the model will now have to be retrained from the 
slice which contains an unlearning request \underline{and}
has the smallest index -- all iterations of training on 
slices that follow it were impacted by the point included in this slice.

To compute the minimum slice index among all 
slices affected by the $K$
unlearning requests, we make the simplifying assumption that multiple unlearning requests are sampled from a uniform distribution \new{$U(1, D)$} \textit{with replacement}. Although this assumption does not hold (the same point would not ask to be unlearned multiple times), we verify numerically that it does not significantly affect our estimate. It is intuitive to see why given that the number
of requests is orders of magnitude smaller than the number of points in the training set.

In Appendix~\ref{appendix:minimum_draw}, we \new{derive the moments of the minimum $X_{min,n}$ of} $n$ draws \new{$X_1, ..., X_{n}$} from an uniform distribution \new{$U(a,b)$}
\new{$\E[\min(X_0, ..., X_n)] = \frac{na + b}{n+1}$.}
\new{This is useful to model the slice of minimum index $r_{min}$ impacted by the batch of unlearning requests.}
We derive the expected \new{cost to be:
\begin{equation}
    \E[C] = \frac{2e'D}{R(R+1)}(\frac{R(R+1)}{2} - \frac{1}{2}(\E[r_{min}^2] - \E[r_{min}])) %
\label{eq: slicing_batch_time_analysis}
\end{equation}}

\new{With $K \gg R$, we have $\E[C] \sim e'D$, which gives no speed-up (but no degradation either).
With $K \ll R$, $\E[C]$ decreases in $\frac{1}{K^2}$ as $K \rightarrow 0$. The case $K = 1$ corresponds to the sequential setting. In that case, we showed a speed-up exists as soon as $R > 1$. Thus there exists a regime, for small values of $K \ll R$, where there is a significant speed-up.} We detail the proof in Appendix~\ref{appendix:time_slicing_batch}.

\section{Implementation Details}

\begin{figure*}[ht]
\centering
{\includegraphics[width=0.8\linewidth]{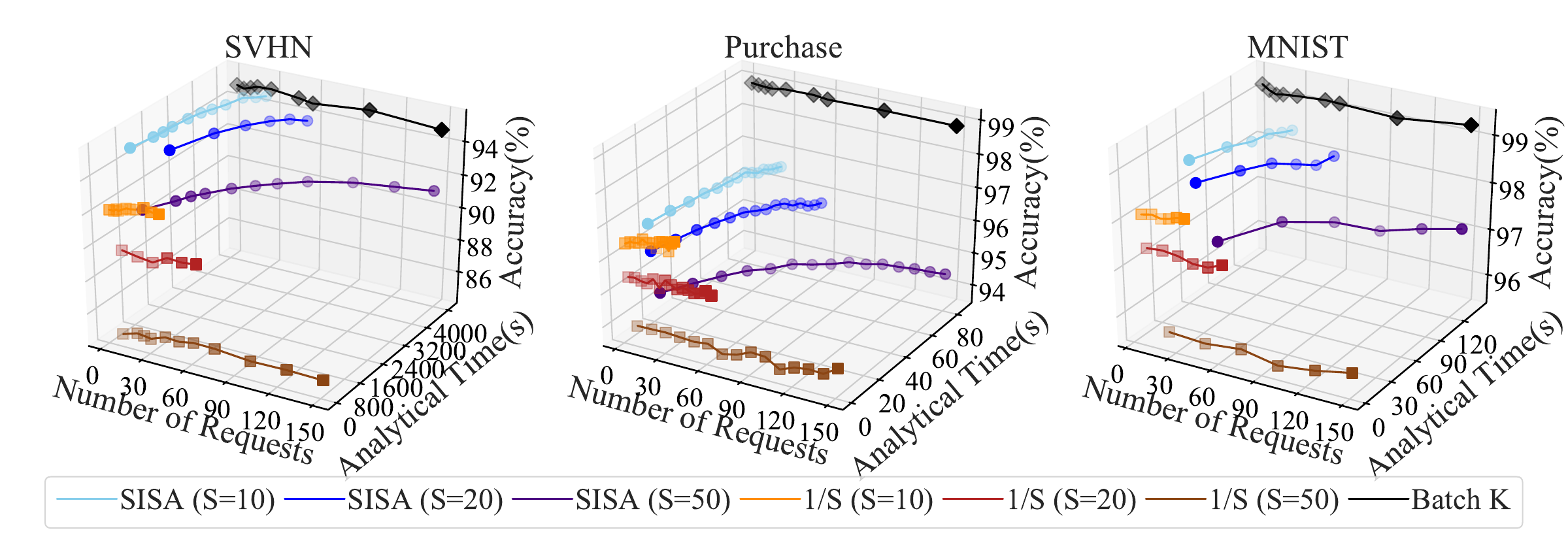}}
\caption{\small \new{We compare the experimental accuracy of \SISA (with different number of shards) with the two baselines on three datasets: SVHN, Purchase, and MNIST. It is clear that \SISA provides higher accuracy than the $\frac{1}{S}$ fraction baseline, along with less retraining time than the batch $K$ baseline especially when the number of unlearning request is small.}\vspace{-5mm}} 
\label{fig:3d_exp}
\end{figure*}

\subsection{Datasets}
\label{datasets}

We provide information about the datasets we used in Table~\ref{tab:dataset}. Note that for the Purchase dataset, we follow a methodology similar to Shokri \etal~\cite[\S 6]{shokri2017membership}; we curated the Purchase dataset by choosing the top 600 most purchased items based on the \texttt{category} attribute. \new{For Mini-Imagenet, we follow the process of Vinyals \etal~\cite{miniimagenet} to create a dataset for supervised classification, not few-shot classification.} 

\begin{table}[H]
    \centering
    \begin{tabular}{l*{3}c}
        \hline
        \textbf{Dataset} & \textbf{Dimensionality} & \textbf{Size} & \textbf{\# Classes} \\
        \hline
        \text{MNIST~\cite{mnist}} & $28\times28$ & 60000 & 10\\
        \text{Purchase~\cite{sakar2019real}} & $600$ & 250000 & 2\\
        \text{SVHN~\cite{netzer2011reading}} & $32\times32\times3$ & 604833 & 10\\
        \text{CIFAR-100~\cite{cifar}} & $32\times32\times3$ & 60000 & 100\\
        \text{Imagenet~\cite{imagenet}} & $224\times224\times3$ & 1281167 & 1000\\
        \text{Mini-Imagenet~\cite{miniimagenet}} & $224\times224\times3$ & 128545 & 100\\

        \hline
    \end{tabular}
    \caption{\small \new{Dataset characteristics.}}
    \label{tab:dataset}
\end{table}

\new{Datasets chosen encapsulate variety in the total number of samples, input dimensionality, samples per class. This allows us to explore a spectrum of {\em task complexities}---the first three are simple while the three remaining are complex. We will highlight the importance of this diversity in later subsections.}

\subsection{Models \& Experimental Setup} 

For simplicity, we use the same model architectures for (a) the baselines and (b) the \SISA scheme. \new{The details are presented in Table~\ref{tab:models}. Observe that we consider a variety of deep neural networks with increasingly more hidden layers as well as varying layer sizes.}

\begin{table}[ht!]
    \centering
    \begin{tabular}{l*{2}c}
        \hline
        \textbf{Dataset} & \textbf{Model Architecture}\\
        \hline
        \text{MNIST~\cite{mnist}} & 2 conv. layers followed by 2 FC layers\\
        \text{Purchase~\cite{sakar2019real}} & 2 FC layers\\
        \text{SVHN~\cite{netzer2011reading}} & Wide ResNet-1-1\\
        \text{CIFAR-100~\cite{cifar}} & ResNet-50\\
        \text{Imagenet~\cite{imagenet}} & ResNet-50\\
        \text{Mini-Imagenet~\cite{miniimagenet}} & ResNet-50\\

        \hline
    \end{tabular}
    \caption{\small \new{Salient features of DNN models used.}}
    \label{tab:models}
\end{table}

We  run  our  experiments  using P100 and T4 Nvidia GPUs, with 12 and 16 GB of dedicated memory, respectively. We use Intel Xeon Silver 4110 CPUs with 8 cores each and 192GB of Ram. The  underlying  OS  is  Ubuntu  18.04.2  LTS  64  bit.We use PyTorch v1.3.1 with CUDA 10.1 and Python 3.6.

\section{Evaluation}
\label{sec:evaluation}

Our evaluation is designed to understand the limitations of \SISA in the scenario where the service provider has no information about the nature of the distribution of the unlearning requests \ie in the uniform setting. In \S~\ref{sec:ulrequests}, we utilize explicit knowledge of this distribution (modeled based on recent public insight from Google~\cite{bertram2019five}) to verify that it improves retraining time. All code (and model checkpoints) are available at \new{\url{https://github.com/cleverhans-lab/machine-unlearning}}. In this section, our experiments tease apart each component of  \SISA. We perform an ablation study to answer the following questions:

\begin{enumerate}
\item What is the impact of sharding on accuracy for varying numbers of unlearning requests? 
\item What is the impact of slicing on accuracy for varying numbers of unlearning requests?
\item Does \SISA improve the retraining time? 
\item \new{Do the findings from above hold for both simple and complex learning tasks?}
\end{enumerate}

\new{We compare our approach against two baselines. They are:
\begin{itemize}
\itemsep0em
\item batch $K$ unlearning requests and retrain the entire model after every $K$ unlearning requests. This is the same to the naive baseline of retraining the entire dataset (without the points to be unlearned) from scratch, in a batch setting.
\item train on a $\frac{1}{S}$ fraction of the data and only retrain when the point to be unlearned falls into this set.
\end{itemize}}
From our analysis, we draw the following insights on the applicability of \SISA in practical settings:

\begin{enumerate}
\item We observe that the sharding component of \SISA induces accuracy degradation as (a) the number of unlearning requests increases, and (b) the number of shards increases (more so for complex tasks). This stems from the decrease in the number of samples per class per shard caused by both (a) and (b) (refer \S~\ref{big}). %
\item We observe that slicing does not induce accuracy degradation so long as the number of epochs required for training are recalibrated (refer \S~\ref{big}).
\item Even in the worst-case scenario (with no knowledge of the distribution of unlearning requests), for a certain number of unlearning requests, a combination of sharding and slicing significantly outperforms the naive baseline. If the number of requests exceeds this threshold, \SISA gracefully degrades to the performance of the baseline. We can analytically obtain this threshold (refer \S~\ref{regime}) based on our theoretical analysis in \S~\ref{sec:time}. 
\item \new{ \SISA has advantages compared to both the batch $K$ baseline, and the $\frac{1}{S}$ fraction baseline in terms of retraining time and accuracy respectively (refer \S~\ref{big}).}
\end{enumerate}

\subsection{The Big Picture}
\label{big}

\begin{figure*}[t]
\centering
\subfloat[{{\small Accuracy vs. Number of epochs for SVHN dataset.}}]{\label{SVHN: accuracy v.s. epochs}\includegraphics[width=0.41\linewidth]{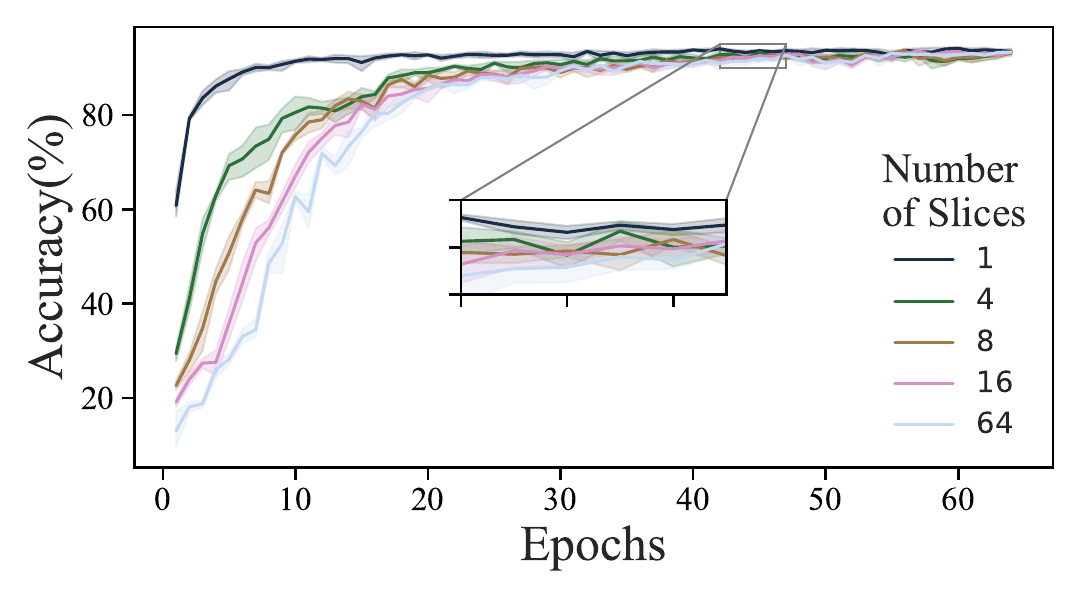}}   
\subfloat[{{\small Accuracy vs. Number of epochs for Purchase dataset.}}]{\label{fig:Purchase: accuracy v.s. epoch}\includegraphics[width=0.41\linewidth]{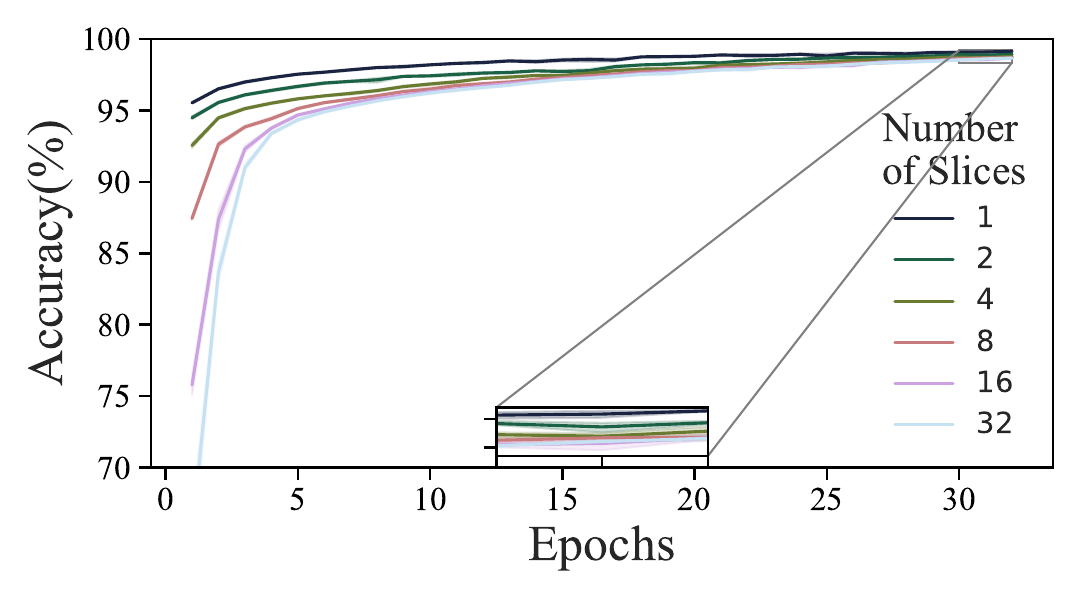}}   
\caption{\small Performance of single model trained with data slicing. We train each model 5 times for each number of slices on the SVHN and Purchase datasets, respectively, and plot the history of validation accuracy and confidence intervals against the number of training epochs. For a small number of epochs, models with more slicing have lower accuracy, due to the fact that they have significantly less amount of data at the beginning. As the number of epochs grows and the accuracy reaches a plateau, the accuracy of models converges. %
} 
\label{fig:slicing}
\vspace{-5mm}
\end{figure*}
To understand the gains, we stress test the approach to understand its benefits for a very large number of shards and a very large number of unlearning requests. \new{In all our experiments (unless mentioned otherwise), \SISA is performed in the batch setting.} 

\subsubsection{Impact of Sharding} As discussed earlier, increasing the number of shards ($S$) increases expected unlearning speed-up (refer \S~\ref{sec:time}) \new{for a bounded number of requests}. However, we wish to understand the impact of sharding on accuracy. To this end, we utilize \SISA for a large number of unlearning requests. \new{Note that the batch $K$ baseline is the same as \SISA with $S=R=1$ in the batch setting.}

\new{From our experiments with simple learning tasks involving the MNIST, SVHN, Purchase datasets (refer Figure~\ref{fig:3d_exp}), we make the following observations: (a) by increasing $S>20$, we observe a more noticeable decrease in accuracy that is greater than 5 percentage points (PPs), and (b) increasing the number of unlearning requests $K>3S$ degrades the retraining time to the batch $K$ baseline case (refer Figures~\ref{fig:svhn_shardvstime} and~\ref{fig:purchase_shardvstime} in Appendix~\ref{appendix:aggregation_strategy}). The former can be attributed to the decreasing volumes of data as the number of shards increases. If the number of shards is greater than 20, we observe that even simple learning tasks (such as those in Figure~\ref{fig:3d_exp}) tend to become more complex (refer \S~\ref{sec:approach}). This phenomenon can also be observed if one increases the number of unlearning requests---after unlearning, each shard has fewer data points.} 

\new{When we compare the accuracy vs. retraining time for \SISA with the 2 baselines, we observe that the batch $K$ baseline has higher accuracy than \SISA, but at the expense of increased retraining time. As noted earlier, this is because this baseline is similar to \SISA with one shard and one slice (ergo losing corresponding speed-ups). The $\frac{1}{S}$ fraction has lower retraining times, but lower accuracy due to the fact that it is trained on a fraction of the entire dataset. While these findings are consistent independently of the task, we discuss the varying impact on accuracy next.}

\begin{figure}[ht!]
\centering
\subfloat[{{\small Imagenet dataset}}]{\label{fig:Imagenet1}{\includegraphics[width=0.5\linewidth]{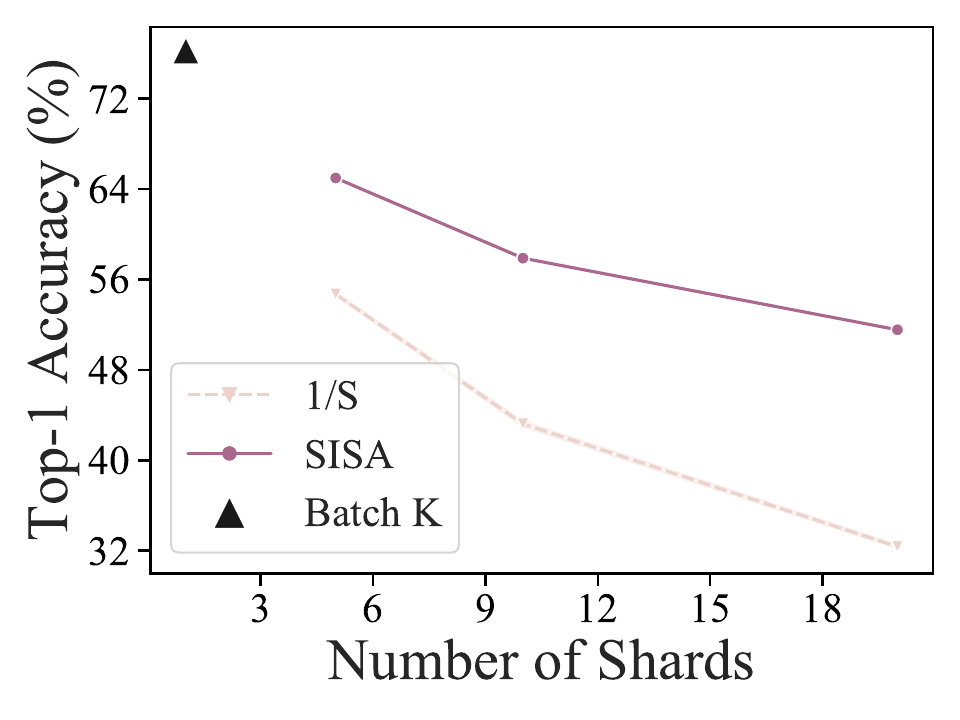}}} 
\subfloat[{{\small Mini-Imagenet dataset}}]{\label{fig:MiniImagenet1}{\includegraphics[width=0.5\linewidth]{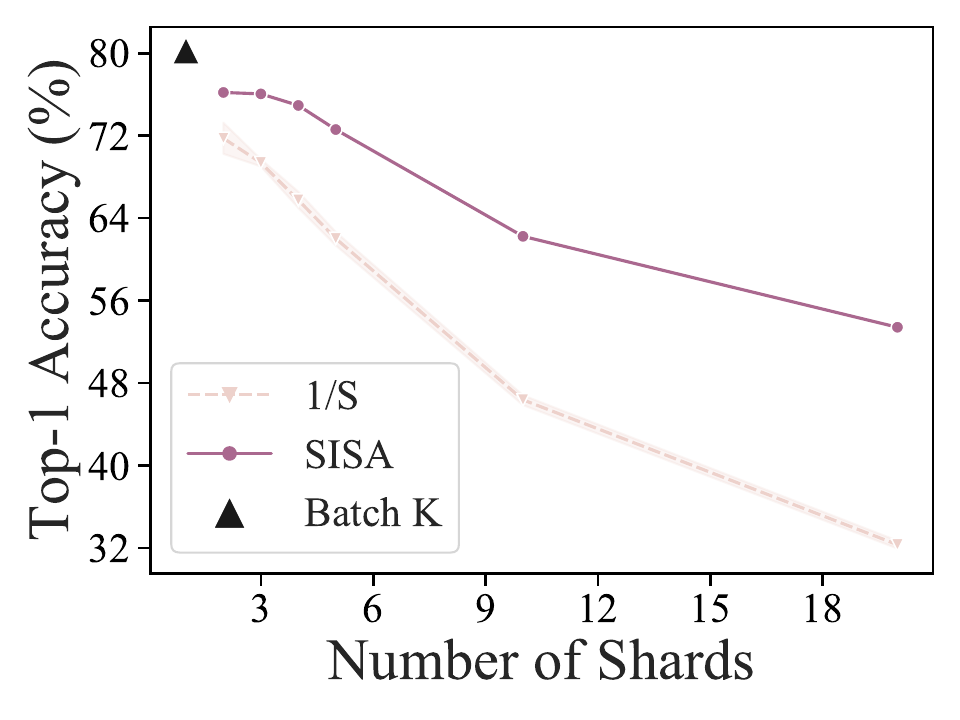}}} 
\caption{\small \new{For complex learning tasks such as those involving Imagenet and Mini-Imagenet, \SISA introduces a larger accuracy gap in comparison to the batch $K$ baseline. However, it is still more performant than the $\frac{1}{S}$ fraction baseline. Each constituent (and baseline) utilized the prediction vector aggregation strategy.}\vspace{-3mm}} 
\label{fig:imagenet}
\end{figure}

\new{Observe that despite having the same benefits over the batch $K$ and $\frac{1}{S}$ fraction baselines, \SISA induces more accuracy degradation for complex tasks (such as Imagenet); from Figure~\ref{fig:imagenet}, observe that \SISA is consistently better than the $\frac{1}{S}$ fraction baseline. However, with label aggregation, the average top-5 accuracy\footnote{The average top-1 accuracy degradation is 18.76 PPs, when the batch $K$ baseline is 76.15\%.} degradation is 16.14 PPs (batch $K$ top-5 accuracy on Imagenet with ResNet-50 is $92.87\%$). To reduce the accuracy gap, we varied the aggregation strategy from label aggregation to prediction vector aggregation (refer \S~\ref{techniques}). From Figure~\ref{fig:imagenet_aggregation} (in Appendix~\ref{appendix:aggregation_strategy}), observe that this provides better accuracy, with average improvements of 1.68 PPs in terms of top-1 accuracy and 4.37 PPs in terms of top-5 accuracy (to reduce the top-5 accuracy gap to 11.77 PPs). We make the same observations on the mini-Imagenet dataset.} %

\new{To validate our belief that the number of samples per class per shard impacts  generalizability of the constituent models, we studied its impact on accuracy. From Figure~\ref{fig:miniimagenet_frac_dataset_accuracy} (in Appendix~\ref{appendix:samples_per_class}), we conclude that the lower number of samples per class per shard (in complex tasks) induces more accuracy degradation. In \S~\ref{acc_gap}, we discuss real world implications of this gap, and how they can be bridged.} 

The key takeaway is that it is essential to ensure each shard has sufficiently many data points to ensure high accuracy at each constituent model.

\subsubsection{Impact of Slicing} From Figure~\ref{fig:slicing}, we observe that slicing \emph{does not have detrimental impact} on model accuracy in comparison to the approach without slicing {\em if} the training time is the same for both approaches. We ensure that training time is the same by setting the number of epochs for slicing based on the calculations in \S~\ref{sec:approach}. \new{Combined with the analysis in \S~\ref{sec:time}, it is clear that slicing {\em reduces} the retraining time so long as the storage overhead for storing the model state after adding a new slice is acceptable (which is linear in the number of slices).}

\begin{figure}[ht!]
\centering
\subfloat[{{\small SVHN dataset}}]{\label{fig:3d_svhn}{\includegraphics[width=0.5\linewidth]{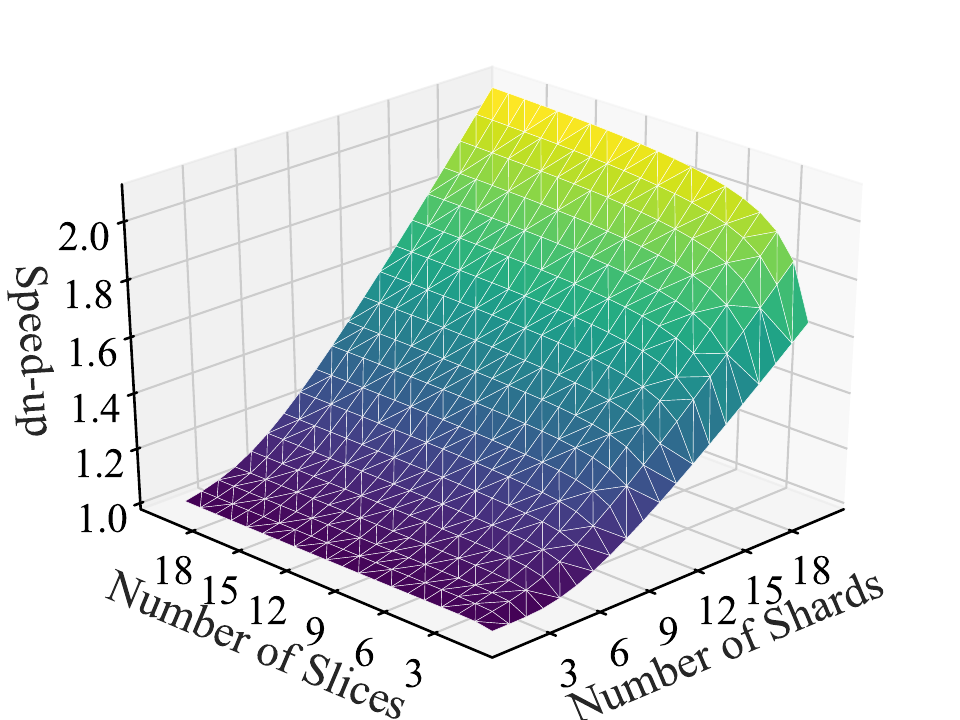}}  } 
\subfloat[{{\small Purchase dataset}}]{\label{fig:3d_purchase}{\includegraphics[width=0.5\linewidth]{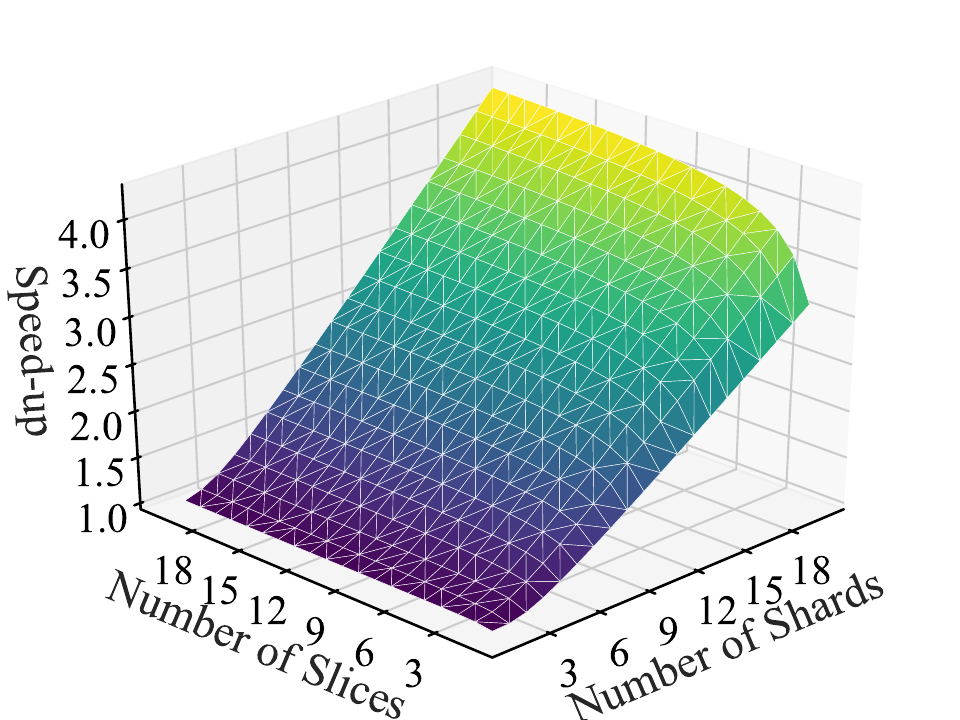}}  } 
\caption{\small \new{Combined speed-up induced by sharding and slicing in the batch setting while there are 0.003\% of the dataset to be unlearned. As the number of shards increases, speed-up increases near proportionally. On the other hand, increasing the number of slices has diminishing returns beyond a few slices.\vspace{-4mm}}} 
\label{fig:3d}
\end{figure}

\subsubsection{Combination of Sharding and Slicing}

\new{From Figure~\ref{fig:3d}, we observe that a combination of sharding and slicing induces the desired speed-up for a fixed number of unlearning requests (0.003\% the size of the corresponding datasets). We utilize these datasets as they have sufficiently many points, resulting in us being able to issue more unlearning requests in the regime where we obtain speed-up (refer \S~\ref{regime}). Observe that the speed-up grows rapidly with an increase in $S$, but increasing $S$ provides marginal gains in this regime. We elaborate upon this further in \S~\ref{regime} and Appendix~\ref{appendix:thought_exp}.}

\subsection{Understanding the Regime}
\label{regime}

The results presented in \S~\ref{big} are exhaustive, and cover a diverse number of shards, slices, unlearning requests, and task complexities. However, not all these configurations are interesting, as some have a detrimental impact on accuracy (as discussed above). \new{For complex learning tasks, better partitioning and aggregation strategies can bridge the accuracy gap, but the findings we present here are generally applicable.} By fixing the number of shards based on our earlier analysis, we can bound the accuracy degradation. However, we wish to understand if there are improvements in retraining time for any number of unlearning requests given this fixed number of shards. Our time analysis in \S~\ref{sec:time} suggests otherwise. Based on this analysis, we plot the retraining time as a function of the number of retraining requests (refer to Figure~\ref{fig:sharding+slicing} in Appendix~\ref{appendix:sharding+slicing}). We observe that for both datasets, the regime where the \SISA approach provides the most retraining benefits is when the number of unlearning requests (as a function of the size of the total dataset) is less than 0.075\% of the dataset. If the number of unlearning requests exceeds this value, then the \SISA approach gracefully degrades to the performance of the batch $K$ baseline. \new{Next, we turn to slicing assuming that the number of shards $S$ is fixed to 20,} and observe that the regime where slicing provides gains is when the number of unlearning requests is less than 0.003\% of the dataset (refer Figure~\ref{fig:sharding+slicing} in Appendix~\ref{appendix:sharding+slicing}). Thus, to extract benefit from both approaches, the ideal number of unlearning requests would be the minimum of the two. \new{Our findings validate that the speed-up exists as long as the number of unlearning requests $K<3S$}. While the regime we provide gains in ($\leq 0.003\%$) may seem very small, recent work by Bertram \etal~\cite{bertram2019five} shows that in practice, the number of unlearning requests (as a function of the size of the total dataset) is much smaller, and is in the order of $10^{-6}$. Additionally, large organizations operate on datasets which are much larger than those in our experiments; the (narrow) regime in which \SISA provides a benefit still provides significant cost reductions.

\subsection{Bridging the Accuracy Gap}
\label{acc_gap}

\begin{figure}[h!]
\centering
{\includegraphics[width=.65\linewidth]{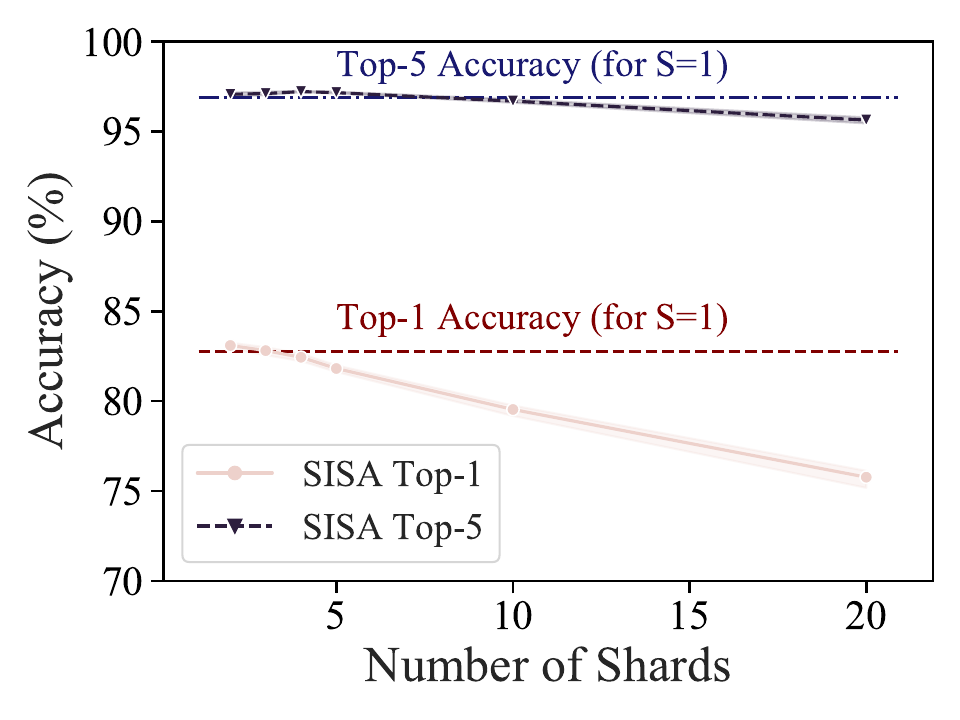}}  
\caption{\small \new{In the setting of transfer learning (from ImageNet to CIFAR-100), we observe a lower accuracy degradation induced by \SISA (with $S > 1$).}\vspace{-3mm}} 
\label{fig:cifar}
\end{figure}

\new{For complex learning tasks in the real-world, the common approach is to utilize a base model trained on public data and utilize transfer learning to customize it towards the task of interest. We replicated such a setup by performing transfer learning using a base model trained on Imagenet (using the ResNet-50 architecture) to the CIFAR-100 dataset. We then perform \SISA and measure the accuracy gap between the baseline ($S=1$) and $S>1$ cases (refer Figure~\ref{fig:cifar}), in terms of both top-1 and top-5 accuracy (the latter is a more representative metric for this complex task). We observe that for this realistic deployment, at $S=10$, the top-1 accuracy gap is reduced to $\sim 4$ PPs, while the top-5 accuracy gap is reduced to $<1$ PP. Additionally in this transfer learning setting, the time analysis for unlearning still holds. Thus, performing transfer learning enables us to decrease the accuracy degradation induced by \SISA on complex tasks without (a) varying the hyperparameters of the constituent models, whilst (b) maintaining constituent model homogeneity}.

\section{Distributional Knowledge}
\label{sec:ulrequests}

In this section, we relax our assumptions and discuss how additional knowledge of the distribution of unlearning requests can be beneficial to the service provider. Specifically, we wish to understand (a) if we can estimate those data points that are more likely to be unlearned than others based on auxiliary information, and (b) how this knowledge can be used a priori to minimize the retraining time and accuracy degradation. 

We believe that an owner's request for unlearning may vary depending on (a) how their data is used and by whom the data is used, (b) the general perception of the surrounding (geographic) population, and (c) incidents related to data misuse etc. For example, machine learning models are not adept at dealing with bias; data owners from those populations who are biased against may wish to request for their data to be erased. By grouping this data, we can further reduce unlearning costs, however, it may also harm fair predictions. Future work should consider these ethical implications. As before, we assume the existence of a data owner $u \in \mathcal{U}$, and the data point generated by $u$ to be $d_u$. We denote the probability of user $u$ requesting to have their data erased as $p(u)$. By aggregating users who are likely to request data erasure into shards of small sizes, intuitively, we would be able to reduce the retraining time.

\begin{figure}[t]
\centering
\includegraphics[width=.8\columnwidth]{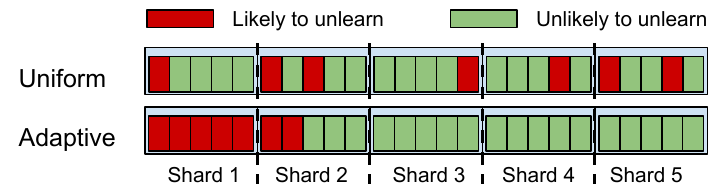}
\caption{\small Example of how a service provider aware of the distribution of unlearning requests may adapt to outperform uniform sharding.}
\label{fig:motivating-example}
\vspace{-5mm}
\end{figure}

To illustrate, consider a population split 
between two groups: the first group $H$ having a high probability $p_H$ of being unlearned
and the second group $L$ having a low probability $p_L$ of being unlearned, with $p_H \gg p_L$. If we follow
the uniform sharding of \S~\ref{sec:approach}, each shard will contain
points from both groups $H$ and $L$. Because points from
$H$ are very likely to be unlearned, and each shard contains at least a few
points from group $H$, it is very likely that all shards will have to be 
unlearned---even if the number of unlearning requests is low. This scenario
is illustrated in Figure~\ref{fig:motivating-example}.  Alternatively, if we know the population will follow such a distribution of unlearning requests, we can adapt our sharding strategy to concentrate
all points from members of group $H$ in relatively few partitions. This strategy ultimately reduces the total number of shards whose models need to be retrained.  We now apply this intuition to a more realistic scenario.

\subsection{Realistic Scenario}

Modeling realistic distributions of unlearning requests is a challenging proposition; prior work in this space is limited. Without data to determine the parameters for a known distribution, such as a Gaussian, or to learn an underlying distribution, we design the following scenario based on insight from the recent work published by Google~\cite{bertram2019five}. Specifically, we propose a scenario where we assume that an organization with access to data records from a large number of data owners operates across various countries, with owners in each region having varied privacy expectations. We assume the existence of $\mathbb{N}$ countries; the dataset $\mathcal{D}$ comprises of per-country datasets $\mathcal{D}_{c}$ for each country $c$.\footnote{Each per-country dataset is conceptually similar to a shard; the distinction is made for easier discussion.} We have $\cap_c \mathcal{D}_c = \varnothing$ and $\cup_c \mathcal{D}_c = \mathcal{D}$. Each data owner in the country $c$ has a fixed probability (denoted $p_c$) for issuing a data erasure request \ie  $\forall d_{u} \in \mathcal{D}_c$, $p(u)=p_c$. Thus, the data owner issuing an unlearning request can be modeled as a Bernoulli trial.

It is important to note that this technique can be generalized to any distribution so long as it is known by the service provider. Specifically, after selecting a distribution $\nu$ that models the unlearning requests from a population $\mathcal{U}$, we randomly sample from this distribution to assign the probability $p(u)$ with which each $u \in \mathcal{U}$ wishes to perform data erasure. Each data point is still a Bernoulli trial; however, the sum \new{$\chi_i$} of these independent Bernoulli trials can be modelled by a Poisson binomial distribution. Armed with this knowledge, we can evaluate the expected number of unlearning requests for this shard $\mathcal{D}_i$, over $n$ trials, as $\E(\chi_i)=n\overline{p}$, where $\overline{p}=\frac{\sum_{u:d_u \in \mathcal{D}_i}p(u)}{|\mathcal{D}_i|}$, and $\E(\chi_i)$ denotes the expectation with which shard $\mathcal{D}_i$ is unlearned. By selecting those users $u \in \mathcal{U}$ and their corresponding data elements $d_u$ to create shard $\mathcal{D}_i$ such that $\E(\chi_i) < C$ for any constant $C \leq 1$, we expect to not have to retrain a model trained using shard $\mathcal{D}_i$. DNNs typically require large data volumes for training; we attempt to create few data shards, with more data in each shard. 

In all experiments we describe in this section, we conceptualize a scenario with $\mathbb{N}=3$ countries -- $c_1,c_2$ and $c_3$, such that $p_{c_1}= 3 \times 10^{-6}$,$p_{c_2}= 3 \times 10^{-5}$, and $p_{c_3}= 6 \times 10^{-6}$. Additionally, $|\mathcal{D}_{c_1}|=0.7717 \times |\mathcal{D}|$ ,$|\mathcal{D}_{c_2}|=0.1001 \times |\mathcal{D}|$ and $|\mathcal{D}_{c_3}|=0.1282 \times |\mathcal{D}|$.

\subsection{Distribution-Aware Sharding} 

\paragraph{Approach} This motivates {\em distribution-aware sharding}, where the service provider can create shards in a way so as to minimize the time required for retraining. We discuss one such approach in Algorithm~\ref{alg:partition}, under the following assumptions: (a) the distribution of unlearning requests is known precisely, and (b) this distribution is relatively constant over a time interval. Recall that each data point $d_u \in \mathcal{D}$ has an associated probability $p(u)$ with which it may be erased. We first sort the data points in the order of their erasure probability, and points to a shard $\mathcal{D}_i$ till the desired value of $\E(\mathcal{D}_i)$ is reached. Once this value is exceeded, we create a new shard $\mathcal{D}_{i+1}$ and restart the procedure with the residual data $\mathcal{D} \setminus \mathcal{D}_i$\footnote{Observe that this strategy holds even when the entire dataset $\mathcal{D}$ is replaced by the dataset for a particular country $\mathcal{D}_c$.}.  By enforcing a uniform cumulative probability of unlearning a across shards, Algorithm~\ref{alg:partition} naturally aggregates the training points that are likely to require unlearning into a fewer shards that are also smaller in size.

\begin{algorithm}
    \caption{Distribution-Aware Sharding}
    \label{alg:partition}
    \hspace*{\algorithmicindent} \textbf{Input:} Dataset $\mathcal{D}$, constant $C$ 
    \begin{algorithmic}[1]
        \Procedure{$ShardData$}{$\mathcal{D}$, $C$}
            \State sort $\{d_{u}\}_{i=1}^{|\mathcal{D}|}$ by $p(u)$
            \State $i \gets 0$
            \State create empty shard $\mathcal{D}_i$
            \For{$j \gets 0$ to $|\mathcal{D}|$}
                \State remove $d_{u}$ with lowest $p(u)$ from $\mathcal{D}$ 
                \State $\mathcal{D}_i = \mathcal{D}_i \cup d_u$
                \If {$\E(\chi_i) \geq C$}
                    \State $\mathcal{D}_i = \mathcal{D}_i \setminus d_u$
                    \State $i \gets i + 1$
                    \State create empty shard $\mathcal{D}_i$
                    \State $\mathcal{D}_i = \mathcal{D}_i \cup d_u$
                \EndIf
            \EndFor
        \EndProcedure
    \end{algorithmic}
\end{algorithm}

\paragraph{Results}

\begin{figure}[t]
\centering
\includegraphics[width=.7\linewidth]{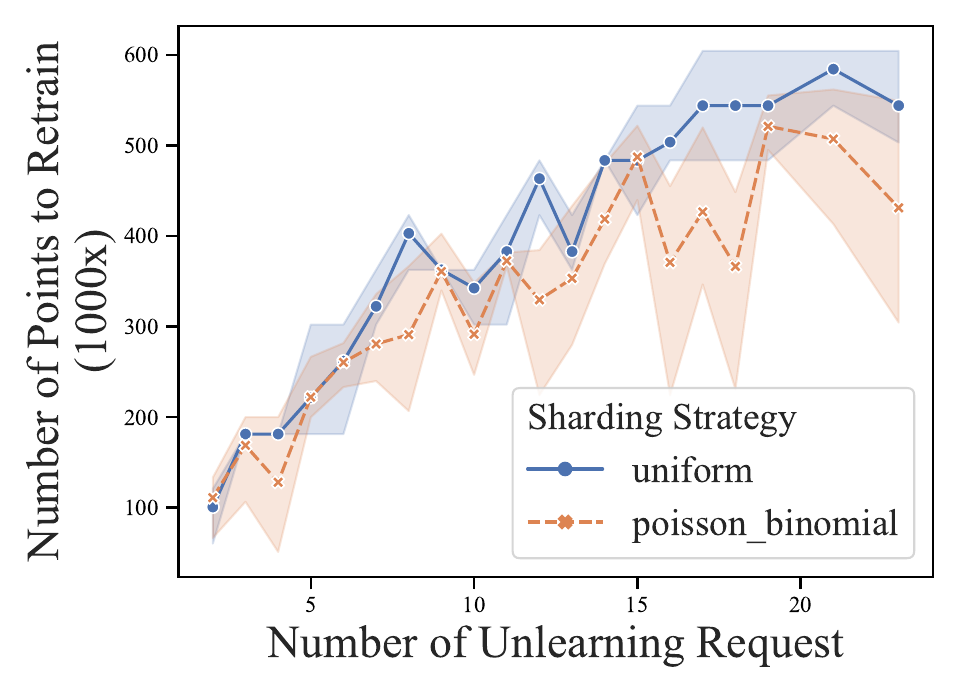}
\caption{\small \# points (variance shaded) of the SVHN dataset that need to be retrained for uniform and distribution-aware sharding where users have varying probability of revoking access to their data.}
\label{fig:realistic-results}
\vspace{-5mm}
\end{figure}

As done for our motivating example, Figure~\ref{fig:realistic-results}
plots the number of points
to be retrained with respect to the number of unlearning requests
for both uniform and distribution-aware sharding. 
In expectation, the distribution-aware strategy decreases the number of points to be 
retrained. Yet, because this strategy creates
shards of unequal size, we also need to evaluate the accuracy of our
predictions aggregated across constituent models. For the parameters specified above, we find that our approach generates 19 shards. We find that the aggregate achieves about 94.4\% prediction accuracy in the regime of unlearning requests we consider, which is one percent point lower than uniform sharding, at 95.7\%. This result means
that distribution-aware sharding incurs a trade-off of accuracy for decreased
unlearning overhead. We leave
to future work the exploration of alternatives to majority voting
aggregation that
would cope with such imbalanced shard sizes. 

\section{Discussion}
\label{sec:discussion}

\vspace{1mm}
\noindent {\em Unlearning in the Absence of Isolation.}
Conceptually, \SISA borrows elements from distributed training and ensemble learning. As discussed earlier, the divide from ensemble learning stems from the fact that each constituent model in \SISA is obtained in isolation. Ensemble learning approaches utilize boosting algorithms~\cite{schapire1999brief}, even for ensembles of neural networks~\cite{schwenk2000boosting}, to enhance accuracy. %

\vspace{1mm}
\noindent  {\em Data Replication.}
Empirical evidence suggests that beyond a certain data volume (\ie shard size), there is performance degradation in each constituent model when datasets are too small, or if the learning task is complex. One way to alleviate this problem is through data replication. However, one must decide which data point is replicated such that the accuracy of the constituent models is increased. This selection is a challenging problem~\cite{settles2009active}. One must also factor in if access to the replicated data point is likely to be revoked; if that is the case, one would intuitively wish to reduce the replication of such a point to limit overhead on unlearning. Understanding these trade-offs is of interest and is future work.

\vspace{1mm}
\noindent  {\em Is All Data Useful?}
Neural networks require large datasets. However, not all of this data is useful~\cite{huang2010active}. As discussed earlier, understanding the importance of each data point towards the final model parameters learned is a challenging problem. A relatively simpler problem is that of {\em core-set selection}, where the objective is to choose a subset of the dataset that will produce a hypothesis that is as performant as one obtained while using the entire dataset~\cite{DBLP:journals/corr/abs-1804-05345,sener2017active}. Core-sets can help reduce the cost of learning. Consequently, they can also improve the cost of unlearning. %

\vspace{1mm}
\noindent  {\em Verified Unlearning.}
We assume that the service provider performs unlearning in an honest manner. Our approach provides an intuitive and provable guarantee under the assumption that the data owner believes the service provider, due to the inherent stochasticity in learning (refer Figure~\ref{fig:unlearning}). 
\new{To increase user confidence, the service provider could release code. One could imagine that authorities relevant to the enforcement of the right to be forgotten could audit the code base to validate the implementation of \SISA. This is sufficient, because of the design of \SISA, to demonstrate that the point to be unlearned would not influence model parameters anymore.}
However, under certain adversarial settings, this trust need not be the case. As stated earlier, there is no way to measure the influence of a data point on the model parameters. Even worse, these models are often proprietary. Thus, understanding if the unlearning procedure can be {\em verified}, similar to approaches in other domains~\cite{tan2017efficient,wahby2017full,setty2012making}, is of merit.

\section{Conclusions}

Our work illustrates how to design learning algorithms that incorporate the need to later unlearn training data. %
We show how simple strategies like \SISA can empower users to expect that their data be completely removed from a model in a timely manner. While our work was primarily motivated by privacy, it is easy to see how unlearning can be a first step towards achieving model governance. We hope this will spur follow-up work on effective ways to patch models upon identifying limitations in datasets used to train them. 

\section*{Acknowledgments}
\noindent We would like to thank the reviewers for their insightful feedback, and Henry Corrigan-Gibbs for his service as the point of contact during the revision process. This work was supported by CIFAR through a Canada CIFAR AI Chair, and by NSERC under the Discovery Program and COHESA strategic research network. We also thank the Vector Institute's sponsors. Varun was supported in part through the following US National Science Foundation grants: CNS-1838733, CNS-1719336, CNS-1647152, CNS-1629833 and CNS-2003129.
\bibliographystyle{IEEEtran}
\bibliography{biblio}
\appendix
\subsection{Simulation of \SISA Time Analysis}
\label{appendix:thought_exp}

\begin{figure}[H]
\centering
\subfloat[{{\small SVHN}}]{\label{fig:log_thought_exp_svhn}{\includegraphics[width=0.5\linewidth]{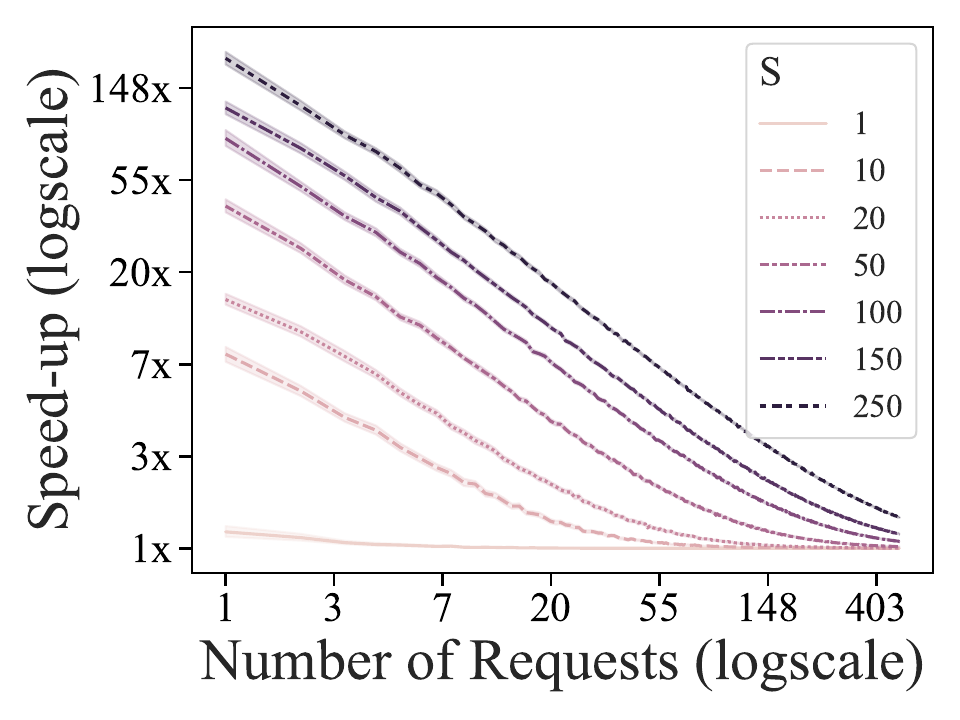}}  } 
\subfloat[{{\small Purchase}}]{\label{fig:log_thought_exp_purchase}{\includegraphics[width=0.5\linewidth]{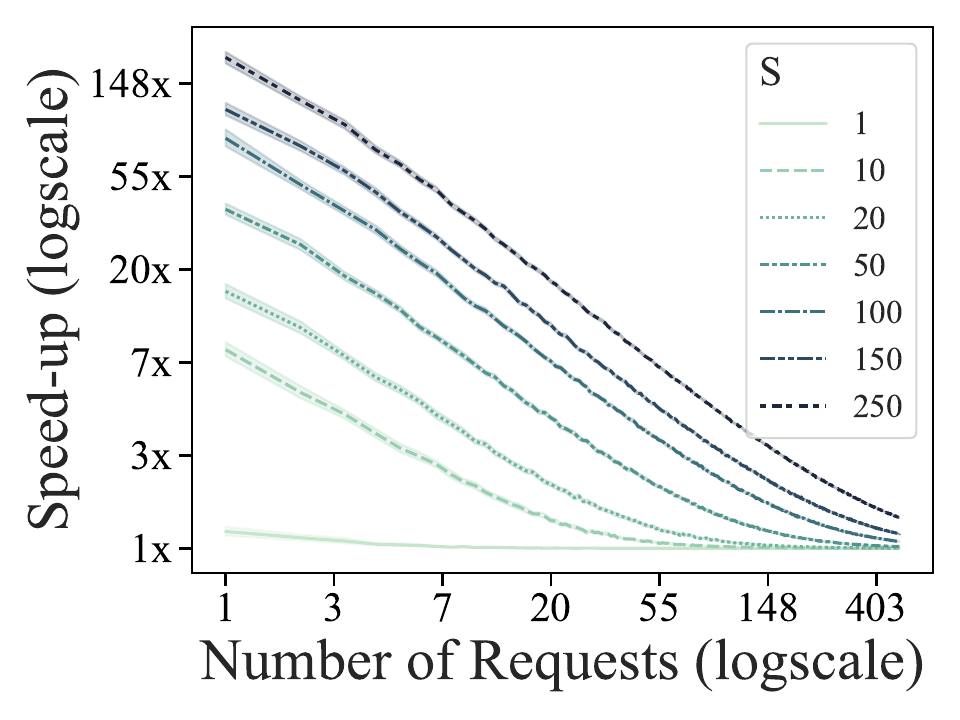}}  } 
\caption{\small \new{This plot shows the relationship between $K$ and unlearning time (which is proportional to amount of data to retrain) where $S$ is shown in the legend and $R$ is set to 20. It is plotted in log-log scale for the ease of visualizing.}\vspace{-4mm}} 
\label{fig:thought_exp}
\end{figure}

\new{To get a more intuitive understanding of unlearning time described in \S~\ref{sec:time}, we randomly generate $K$ unlearning requests. We then compute the amount of data that needs to be retrained by determining the shard and slice each unlearning request maps to. We then deduce the number of samples that need to be retrained on, to achieve unlearning through \SISA. By varying $K$ between 1 and 500, we visualize the speed-up achieved by \SISA as a function of the number of unlearning requests made. We repeat the experiment 100 times to obtain variance. The results are plotted in Figure~\ref{fig:thought_exp}.}

\subsection{\new{Individual Contributions due to Slicing and Sharding}}
\label{appendix:sharding+slicing}

\begin{figure*}[ht]
\centering
\subfloat[{{\small Impact of sharding on the number of points to retrain(SVHN)}}]{\label{fig:svhn_shardvstime}{\includegraphics[width=0.35\linewidth]{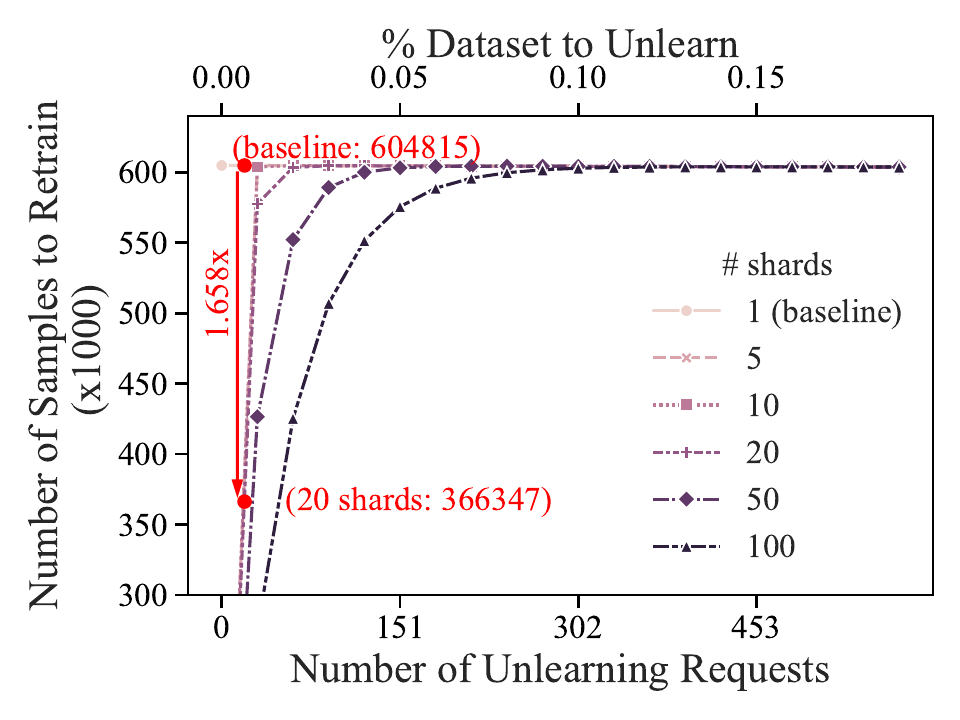}}  } 
\subfloat[{{\small Impact of slicing on the number of points to retrain(SVHN)}}]{\label{fig:svhn_slicevstime}{\includegraphics[width=0.35\linewidth]{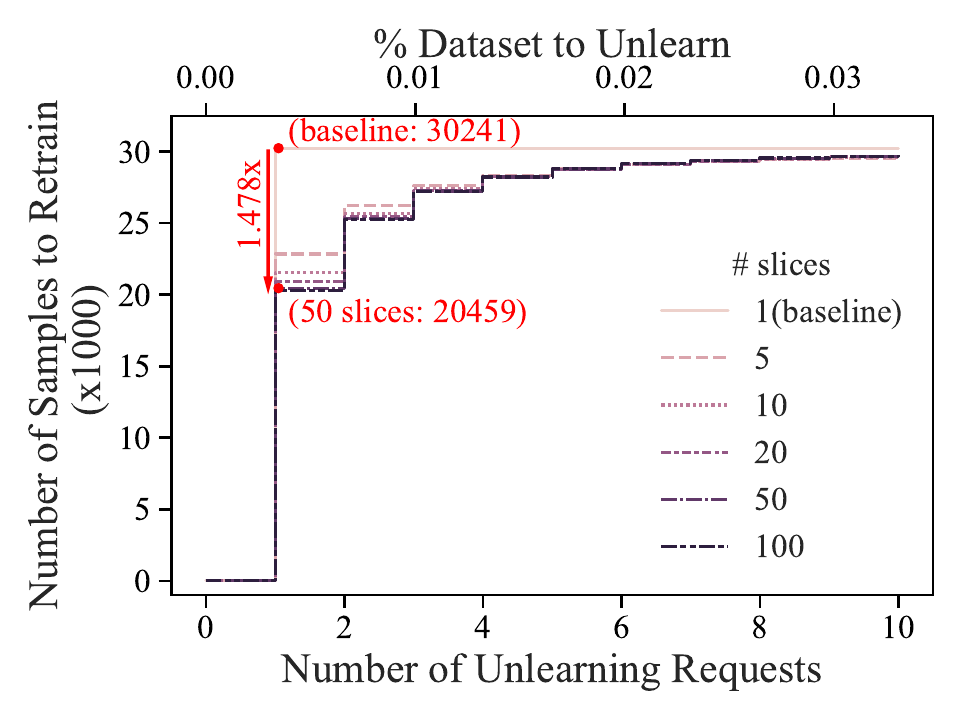}}  } 

\subfloat[{{\small Impact of sharding on the number of points to retrain(Purchase)}}]{\label{fig:purchase_shardvstime}{\includegraphics[width=0.35\linewidth]{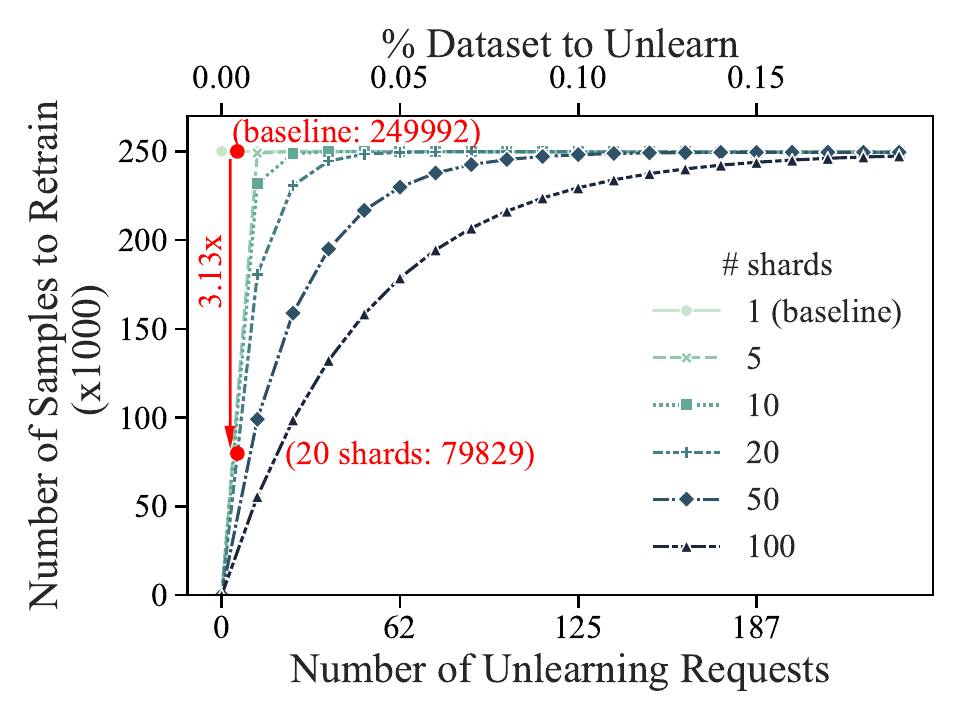}}  } 
\subfloat[{{\small Impact of slicing on the number of points to retrain(Purchase)}}]{\label{fig:purchase_slicevstime}{\includegraphics[width=0.35\linewidth]{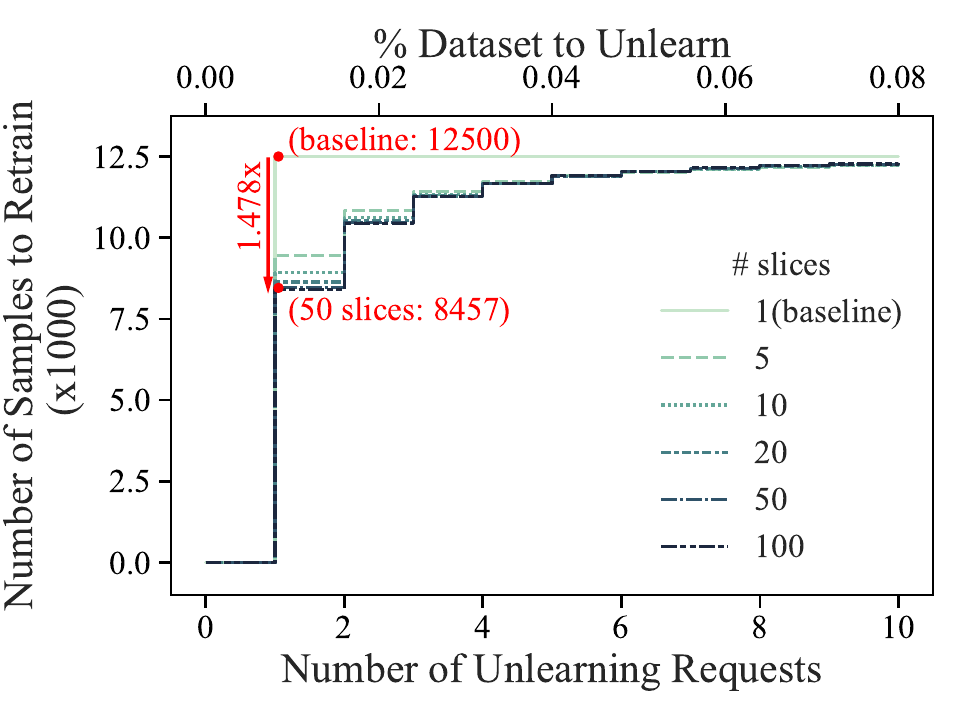}}  } 

\caption{\small Impact of sharding and slicing on retraining time \new{in a batch setting}, as measured by the changes induced in the number of points needed for retraining (which is a proxy for retraining time). Observe that below a particular number of unlearning requests, both sharding and slicing provide noticeable improvements. Afterward, both gracefully degrade to the performance of the naive baseline.} 
\label{fig:sharding+slicing}
\vspace{-5mm}
\end{figure*}

\new{In Equations~\ref{eq: sharding_batch_time_analysis} and \ref{eq: slicing_batch_time_analysis}, we present the unlearning cost (\ie number of points needed to be retrained) as functions of number of unlearning requests, slices, and shards. We plotted the speed-up induced by SISA in Figure~\ref{fig:3d}, but the number of unlearning requests is set to a constant for ease of visualization. Therefore, Figure~\ref{fig:sharding+slicing} is plotted to show the effect of all three variables.}

\subsection{Costs Associated With Storage}
\label{app:storage}

\new{The slicing introduced by \SISA is trading disk storage for unlearning speed-up. This is supported by the fact that the the cost of GPU accelerators is more than the cost of storage disks. For instance, storage costs \$0.026/Gb/Month on Google Cloud, \$0.023/Gb/Month on Amazon Web Services, and \$0.058/GB per month on Azure at the time of writing. Instead, renting the cheapest GPUs starts at \$0.25/hour on Google Cloud, \$0.526/hour on Amazon Web Services, and \$0.90/hour on Azure. 
To limit usage of GPU accelerators, it is already a common practice to regularly checkpoint models during training. Hence, we believe the overhead from slicing will not be a drawback: in most cases, it will suffice to change the order in which data is presented during training because checkpointing is already in place.}	
\subsection{Sequential Time Analysis of Sharding}
\label{appendix:time_sharding_sequential}

\noindent{\em Proof:} \\
\noindent\emph{1. Assumption:} At each step and for all shards, the probability that an unlearning request affects that specific shard is approximately equal to $\frac{1}{S}$. The intuition is as follows: if many points from a specific shard are deleted as unlearning occurs, the number of such (unlearnable) points decreases and they are therefore less likely to be deleted. Conversely, if few points from that shard are deleted, the proportion of those points increases as points from other shards are deleted. Thus, they become more likely to be deleted. 

\noindent\emph{2. Intuition:} The size of the shard that is affected by the first request is always $\frac{N}{S}$. For the second request, it can be either $\frac{N}{S}$ with probability $\left(1 - \frac{1}{S}\right)$ if the request does not affect the same shard or $\left(\frac{N}{S} - 1\right)$ with probability $\frac{1}{S}$ if it does. For the third request it can be $\frac{N}{S}$, $\left(\frac{N}{S} - 1\right)$, or $\left(\frac{N}{S} - 2\right)$. Note that there are two ways to get $\left(\frac{N}{S} - 1\right)$: either from a shard that had $\left(\frac{N}{S} - 1\right)$ point before the previous request and that was not affected by it, or from a shard that had $\frac{N}{S}$ points before the previous request and that was affected by it.

\noindent\emph{3. Size of the retraining set:} To model this behavior, we define the event $E_{i,j}$ as
the $i^{th}$ request received landing on shard $s$ containing $\frac{N}{S} - j$ points, with $j \in \{0,\dots,i - 1\}$. The associated cost is $\frac{N}{S} - j - 1$.

\noindent\emph{4. Associated probability:} The probability of $E_{i,j}$ given a configuration of the $j$ requests, \ie which specific subset of the $i-1$ requests corresponds to those that landed on $s$, is $\left(\frac{1}{S}\right)^j\left(1 - \frac{1}{S}\right)^{i-1-j}$. The first term of the product means that shard $s$ was affected $j$ times, and the second term means that another shard (but not $s$) was affected $i - 1 - j$ times. However, there are ${i - 1 \choose j}$ possible configurations of the $j$ requests that landed on shard $s$. Thus the probability of $E_{i,j}$ is ${i-1 \choose j}\left(\frac{1}{S}\right)^j\left(1-\frac{1}{S}\right)^{i-j-1}$.

\noindent\emph{5. Expected cost:} The expected cost of the $i^{th}$ unlearning request can be decomposed on the family of events $(E_{i,j})_j$ (with only $j$ varying between $0$ and $i-1$) that partitions the universe of possible outcomes:
\begin{equation}\mathbb{E}[C_i] =\sum_{j=0}^{i-1}{{i-1 \choose j}\left(\frac{1}{S}\right)^j\left(1-\frac{1}{S}\right)^{i-j-1} \left(\frac{N}{S} - 1 - j\right)}\end{equation}

To obtain the total cost, we sum the costs of all unlearning requests, which is to say we sum over $i$ between $1$ and $K$.

{\small
\begin{equation}\begin{aligned}
\mathbb{E}[C] &=\sum_{i=1}^{K}\sum_{j=0}^{i-1}{{i-1 \choose j}\left(\frac{1}{S}\right)^j\left(1-\frac{1}{S}\right)^{i-j-1} \left(\frac{N}{S} - 1 - j\right)}\\
&=\sum_{i=1}^{K}{\left(\frac{N}{S} - 1\right)\sum_{j=0}^{i-1}{{i-1 \choose j}\left(\frac{1}{S}\right)^j\left(1-\frac{1}{S}\right)^{i-j-1}}}\\
&-\sum_{i=1}^{K}\sum_{j=0}^{i-1}{j{i-1 \choose j}\left(\frac{1}{S}\right)^j\left(1-\frac{1}{S}\right)^{i-j-1}}
\end{aligned}\end{equation}
}

We can use the fact that $j{i-1 \choose j} = (i-1){i-2 \choose j-1}$ and apply the binomial theorem to both inner sums after reindexing the second inner sum.

{\small
\begin{equation}\begin{aligned}
\mathbb{E}[C] &= \left(\frac{N}{S} - 1\right)K - \sum_{i=1}^{K}{\frac{i-1}{S}}\\
\end{aligned}\end{equation}
}

\subsection{Batched Time Analysis of Sharding}
\label{appendix:time_sharding_batch}

\noindent{\em Proof:} Let $S$ denote the number of shards and $K$ the number of points in the batch.
Let $(\mathfrak{s}_i)_{i \in \{1,\dots,K\}}$ be random variables that give the index of the shard impacted by each point in the batch. We assume that those variables are i.i.d. and that:
\[\forall i \in \{1,\dots,K\}, \mathfrak{s}_i \sim U(0, S)\]
We can define $(\mathfrak{h}_j)_{j \in \{1,\dots,S\}}$ which are Bernoulli random variables whose value is 1 when shard $j$ is impacted.
We have:

{\small\[\mathfrak{h}_j = 0 \Longleftrightarrow \forall i \in \{1,\dots,K\}, \mathfrak{s}_i \ne j\]}

Thus $\mathbb{P}(\mathfrak{h}_j = 0) = \left(1 - \frac{1}{S}\right)^K$. We define the total cost $C$ of retraining as the number of points that need to be retrained while processing the batch as $C = \sum_{j=1}^S\mathfrak{h}_j|\mathcal{D}_j|$

To obtain $|\mathcal{D}_j|$, we define $(\mathfrak{u}_j)_{j \in \{1,\dots,K\}}$, the random variables that count the number of times each shard is affected. These variables count the number of successes in a repetition of independent Bernoulli experiments, namely counting the number of times $\mathfrak{s}_i = j$, when $i$ varies from $1$ to $K$. Thus:
{\small\[\forall j \in \{1,\dots,S\}, \mathfrak{u}_j \sim B\left(K, \dfrac{1}{S}\right)\]}

{\small\begin{equation}\begin{aligned}
C &= \sum_{j=1}^S{\mathfrak{h}_j\left(\frac{N}{S} - \mathfrak{u}_j\right)}
&= \sum_{j=1}^S{\left(\frac{N\mathfrak{h}_j}{S} - \mathfrak{u}_j\mathfrak{h}_j\right)}
\end{aligned}\end{equation}}
By construction,
{\small\[\mathfrak{h}_j = 0 \Longleftrightarrow \mathfrak{u}_j = 0\]}
Thus $\mathfrak{u}_j\mathfrak{h}_j = \mathfrak{u}_j$ and:
{\small\[C = \sum_{j=1}^S{\left(\frac{N\mathfrak{h}_j}{S} - \mathfrak{u}_j\right)}\]}
Using the linearity of the expected value and the expected values of Bernouilli and binomial random variables,
{\small\begin{equation}\begin{aligned}
\mathbb{E}[C] &=& \sum_{j=1}^S{\left(\frac{N}{S}\left(1-\left(1-\frac{1}{S}\right)^K\right) - \frac{K}{S}\right)}\\
&=& N\left(1-\left(1-\frac{1}{S}\right)^K\right) -K
\end{aligned}\end{equation}}
	
\subsection{Sequential Time Analysis of Slicing}
\label{appendix:time_slicing_sequential}

\noindent{\em Proof:} When a model is trained on an entire shard (\ie without slicing) of size $D=\frac{N}{S}$ for $e'$ epochs, the number of samples seen by the training algorithm is proportional to $e'D$.
Recall from \S~\ref{sec:approach} that we modified the number of epochs $e$ when slicing is applied (refer to equation~\ref{eq:slice_epoch}).
For each slice indexed $r$, we use data from the first $r$ slices (\ie $\frac{rD}{R}$ samples), training the model for $\frac{2e'}{R+1}$ epochs. Therefore, if an unlearning request hits the $r^{th}$ slice, we need to retrain the model from the $r^{th}$ slice to the $R^{th}$ slice, leading to the following retraining cost (\ie number of samples):

{\small\begin{equation}\begin{aligned}
    C &= \sum_{j = r}^R{\frac{2e'}{R+1}\frac{jD}{R}}
    &= \frac{2e'D}{R(R+1)}\left(\frac{R(R+1)}{2} - \frac{r(r-1)}{2}\right)
\end{aligned}\end{equation}}

We model the index of a slice hit by an unlearning request by the random variable $\mathfrak{r} \sim U(\{1,\dots,R\})$.
The expected cost can be expressed as:
{\small\begin{equation}\begin{aligned}
    \mathbb{E}[C] 
    &= \frac{2e'D}{R(R+1)}\left(\frac{R(R+1)}{2} - \frac{1}{2}\left(\mathbb{E}[\mathfrak{r}^2] - \mathbb{E}[\mathfrak{r}]\right))\right)
\end{aligned}\end{equation}}
We can compute the two first moments of $\mathfrak{r}$:

{\small\[\begin{array}{lclcl}
    \mathbb{E}[\mathfrak{r}] &=& \sum_{k=1}^R{k\mathbb{P}(\mathfrak{r}=k)} &=& \frac{R+1}{2}\\
    \mathbb{E}[\mathfrak{r}^2] &=& \sum_{k=1}^R{k^2\mathbb{P}(\mathfrak{r}=k)} &=& \frac{(R+1)(2R+1)}{6}\\
\end{array}\]}

And plug them into the expected cost:
{\small\begin{equation}\begin{aligned}
    \mathbb{E}[C] &= \frac{e'D}{R}\left(R - \frac{2R + 1}{6} + \frac{1}{2}\right)
    &= e'D\left(\frac{2}{3} + \frac{1}{3R}\right)\\
\end{aligned}\end{equation}}
\new{Note that for $R > 20$, the speed-up starts to plateau and any increase in $R$ does not provide a significant speed-up gain.}

\subsection{Moments of the Minimum of Draws from a Uniform Distribution}
\label{appendix:minimum_draw}

Let $X_1, ..., X_n$ denote the $n$ draws we make from a uniform distribution $U([a,b])$.
We would like to compute the expectation of the minimum of these draws, denoted as $X_{min,n}=\min_{i\in\{1,\dots,n\}}(X_i)$. 

\noindent{\em Proof:} Our proof follows material found online~\cite{math}.
First recall that the cumulative distribution function of any $X_i$ is $F_{X_i} = \frac{x-a}{b-a}\mathbb{1}_{[a,b]} + \mathbb{1}_{[b,+\infty)}$
We then compute the CDF of $X_{min,n}$:
{\small
\begin{equation}
\begin{aligned}
    F_{X_{min,n}}(x) &= \mathbb{P}(X_{min,n} \leq x)\\
    &= 1-\mathbb{P}(X_{min,n} > x)\\
    &= 1 - \mathbb{P}\left(\bigcap_{i=1}^n{(X_i > x)}\right)\\
    &= 1-\prod_{i=1}^n{\mathbb{P}(X_i > x)}\\
    &= \left(1-\prod_{i=1}^n{\left(1 - \frac{x-a}{b-a}\right)}\right)\mathbb{1}_{[a,b]} + \mathbb{1}_{[b,+\infty)}\\
    &= \left(1-\left(\frac{b-x}{b-a}\right)^n\right)\mathbb{1}_{[a,b]} + \mathbb{1}_{[b,+\infty)}\\
\end{aligned}
\end{equation}}
where the antepenultimate line holds because the draws are independent. We now take the derivative and obtain the density function:
{\small\begin{equation}f_{X_{min,n}}(x) = \frac{n}{b-a} \left(\frac{b-x}{b-a}\right)^{n-1}\mathbb{1}_{[a,b]}\end{equation}}

We compute the first moment of $X_{min,n}$ by using an integration by part:
{\small\begin{equation}
    \label{eq:min}
    \mathbb{E}[X_{min,n}] = \int_{-\infty}^{+\infty} xf_{X_{min,n}}(x) \mathrm{d}x = \frac{na + b}{n+1}
\end{equation}}

Similarly, we can compute the second moment by using two integrations by part (or the first moment of $X_{min,n+1}$):
{\small\begin{equation}
\begin{aligned}
    \label{eq:min_square}
    \mathbb{E}[X_{min,n}^2] &= \int_{-\infty}^{+\infty}{x^2f_{X_{min,n}}(x)\mathrm{d}x}\\
    &= a^2 + \frac{2(b-a)}{n+1}\frac{(n+1)a+b}{n+2}
\end{aligned}
\end{equation}}

\subsection{Batched Time Analysis of Slicing}
\label{appendix:time_slicing_batch}

\noindent{\em Proof:} In the batch setting, we retrain all the slices between the slice $\mathfrak{r}_{min}$ having the minimal index that has been hit after $K$ requests and the $R^{th}$ slice. Since the indices $(\mathfrak{r}_i)_{i \in \{1,\dots,K\}} \sim U(\{1,\dots,R\})$ i.i.d. (we assume the requests to be independent), we can use results of previous sections to compute the moments of $\mathfrak{r}_{min}$. The expected cost becomes:
{\small\begin{equation}\begin{aligned}
    \mathbb{E}[C] 
    &= \frac{2e'D}{R(R+1)}\left(\frac{R(R+1)}{2}\right.\\
    &- \left.\frac{1}{2}\left(1 + \frac{2(R-1)}{K+1}\frac{(K+1)+R}{K+2} - \frac{K+R}{K+1}\right)\right)
\end{aligned}\end{equation}}
	
\subsection{Lone Shard Baseline Time Analysis}
\label{appendix:time_baseline}

\noindent{\bf Definition:} A lone shard is a model trained on a $\frac{1}{S}$ fraction of the dataset. The remainder of data is not used for training.

\subsubsection{Sequential Setting}
\label{appendix:seq_lon_shard}

\noindent

\noindent\emph{1. Assumption:} The assumptions made in Appendix~\ref{appendix:time_sharding_sequential} are valid here, though we only have one shard of initial size $\frac{N}{S}$. The probability of it being impacted is approximately equal to $\frac{1}{S}$.

\noindent\emph{2. Size of the retraining set:} We can develop a reasoning very similar to Appendix~\ref{appendix:time_sharding_sequential}. At each step, two cases are possible. Either we affect a shard, the only shard we have. This corresponds to the event $E_{i,j}$ of Appendix~\ref{appendix:time_sharding_sequential} if the shards has already been affected $j$ times, or we affect no shard with cost $0$. We call this event $Z_i$.

\noindent\emph{3. Associated probabilities:}
The probability of $Z_i$, since we have only one shard, is $1 - \frac{1}{S}$. Notice that in Appendix~\ref{appendix:time_sharding_sequential} this event had zero probability.
The probability of $E_{i,j}$ is $\frac{1}{S}{i - 1 \choose j}\left(\frac{1}{S}\right)^{j}\left(\frac{1}{S}\right)^{i - 1 - j}$. The factor $\frac{1}{S}$ accounts for the fact that request $i$ affects a shard with probability $\frac{1}{S}$, the rest of the formula is similar to the one in Appendix~\ref{appendix:time_sharding_sequential}.

\noindent\emph{4. Expected cost:}
We can easily show that we obtain a formula for the expected cost similar to the one in Appendix~\ref{appendix:time_sharding_sequential} but with a $\frac{1}{S}$ factor:
\begin{equation}\mathbb{E}[C] = \frac{1}{S}\left(\frac{N}{S} + \frac{1}{2S}-1\right)K - \frac{K^2}{2S^2}\end{equation}
Thus the lone shard baseline provides a $S\times$ speed up w.r.t. \SISA. However, this fact should not discourage the use of \SISA since the lone shard baseline will perform poorly in terms of accuracy on complex learning tasks.

\subsubsection{Batched Setting}

Let $K$ denote the batch size. We model whether the $i^{th}$ request of the batch affects the training set (or not) by a Bernoulli random variable $\mathfrak{h}_i \sim B(\frac{1}{S})$ i.i.d. We define $\mathfrak{s}_K = \sum_i{\mathfrak{h}_i}$ the number of times the shard is affected for the batch. By construction, $\mathfrak{s}_K \sim B(K,\frac{1}{S})$.
The number of points to retrain when the batch is processed is simply the total number of points in the training set minus the number of times the shard is affected: $C = \frac{N}{S} - \mathfrak{s}_K$. Thus:
{\small\begin{equation}
\mathbb{E}[C] = \frac{N-K}{S}
\end{equation}}

Recall that the batched cost of \SISA is:
{\small\begin{equation}
\mathbb{E}[C] = N\left(1-\left(1-\dfrac{1}{S}\right)^K\right) -K
\end{equation}}

For $K \ll N$, we roughly have a cost of $N(1-\exp({\frac{-K}{\tau})})$ where $\tau = (-\ln({1-\frac{1}{S}}))^{-1}$ for \SISA.

Thus for small enough $K$, there {\em might exist} a regime where \SISA outperforms the lone shard baseline. Determining a usable value of $K$ in that regime is the challenge -- $K$ can not be less than 1. Note that $K=1$ is exactly the first step of the sequential setting: the lone shard baseline provides a $S\times$ speed up w.r.t. \SISA (refer \S~\ref{appendix:seq_lon_shard}). It turns out this regime is impractical. Therefore, for small values of $K$, the lone shard baseline outperforms \SISA with a speed-up of at least $S \times$. Once again, those findings must be considered along with their impact on accuracy, and are meaningless by themselves.	
\subsection{Impact of aggregation strategy}
\label{appendix:aggregation_strategy}

\begin{figure}[H]
\centering
\subfloat[{{\small SVHN}}]{\label{fig:svhn_aggregation}{\includegraphics[width=0.5\linewidth]{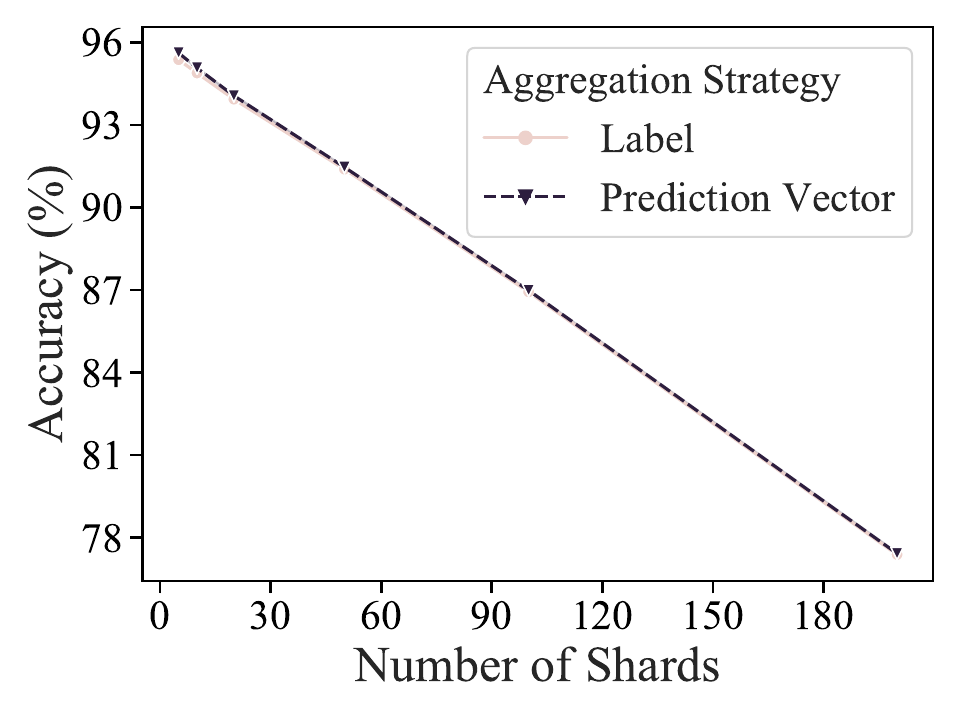}}  } 
\subfloat[{{\small Purchase}}]{\label{fig:purchase_aggregation}{\includegraphics[width=0.5\linewidth]{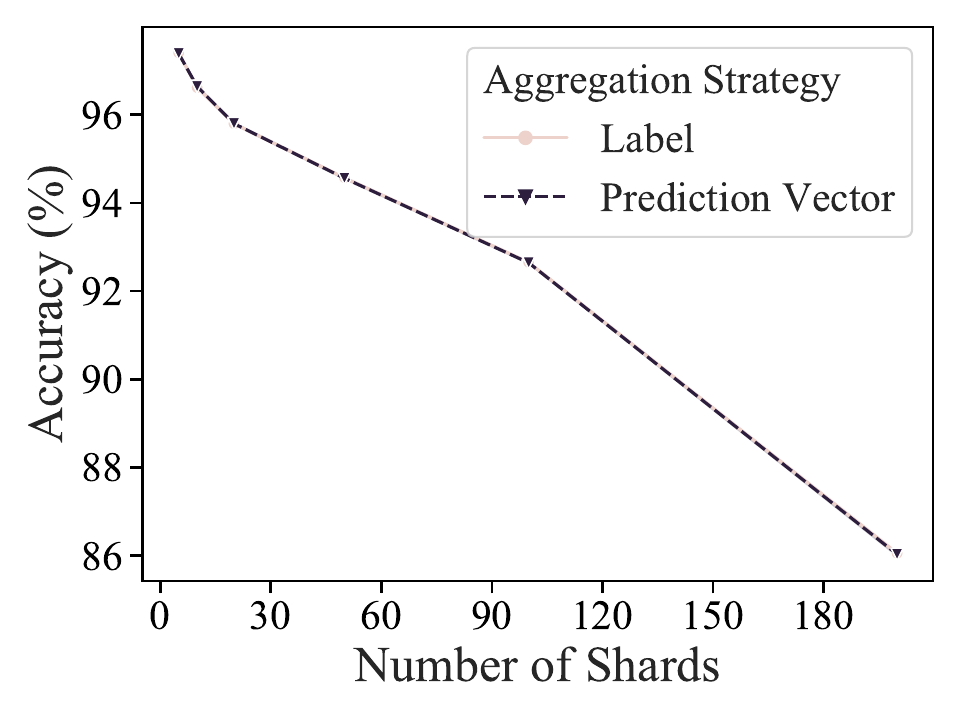}}  } 
\caption{\small \new{We explore the effect of aggregating labels versus aggregation prediction vectors, on Purchase and SVHN. It can be seen that on these easy datasets, changing aggregation strategy does not influence performance of the models significantly.}\vspace{-5mm}} 
\label{fig:aggregation_strategy1}
\end{figure}

\begin{figure}[H]
\centering
\subfloat[{{\small Imagenet}}]{\label{fig:imagenet_aggregation}{\includegraphics[width=0.5\linewidth]{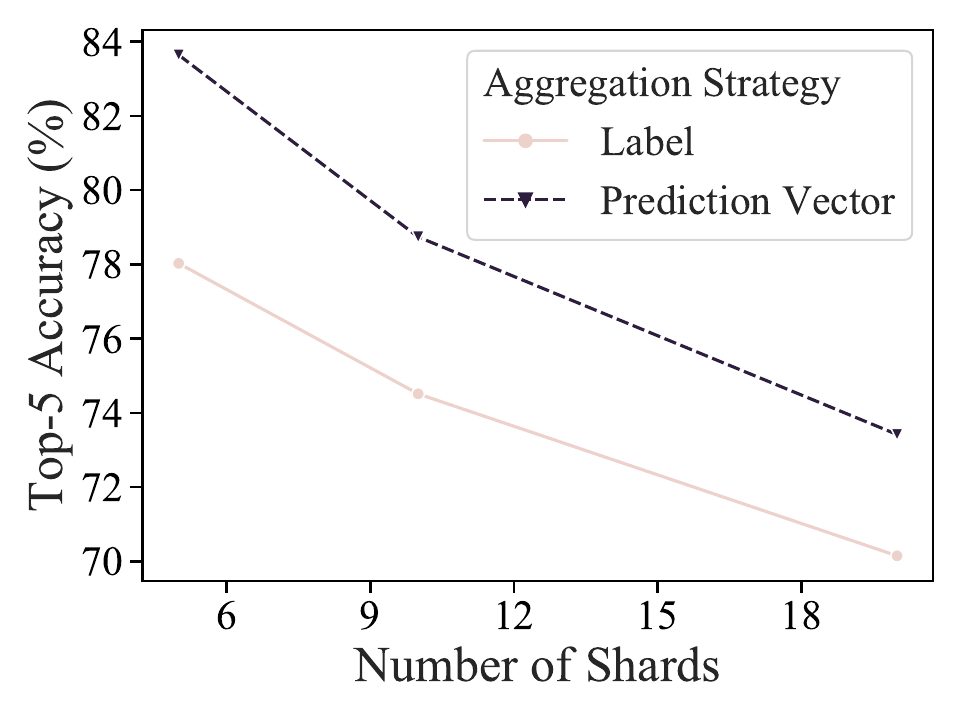}}  } 
\subfloat[{{\small Mini-Imagenet}}]{\label{fig:miniimagenet_aggregation}{\includegraphics[width=0.5\linewidth]{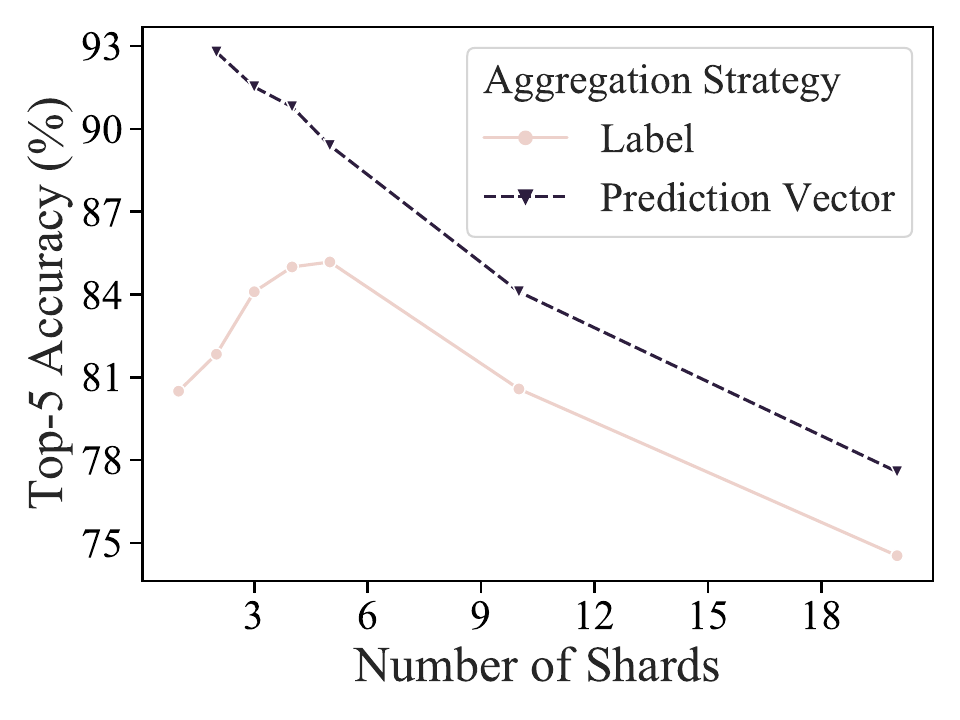}}  } 
\caption{\small \new{We explore the effect of aggregating labels versus aggregation prediction vectors, on Mini-ImageNet and ImageNet. It can be seen that on these hard tasks such as classifying high-resolution images, a good aggregation strategy is able to help recover more accuracy.}\vspace{-5mm}} 
\label{fig:aggregation_strategy2}
\end{figure}

\new{Due to the nature of SISA, we need to aggregate the predictions of different models. Here we tested 2 aggregation strategies on 4 datasets respectively, the results can be found in Figures~\ref{fig:aggregation_strategy1} and~\ref{fig:aggregation_strategy2}.}

\subsection{Impact of number of samples per class on learnability}
\label{appendix:samples_per_class}
\begin{figure}[h]
\centering
{\includegraphics[width=.6\linewidth]{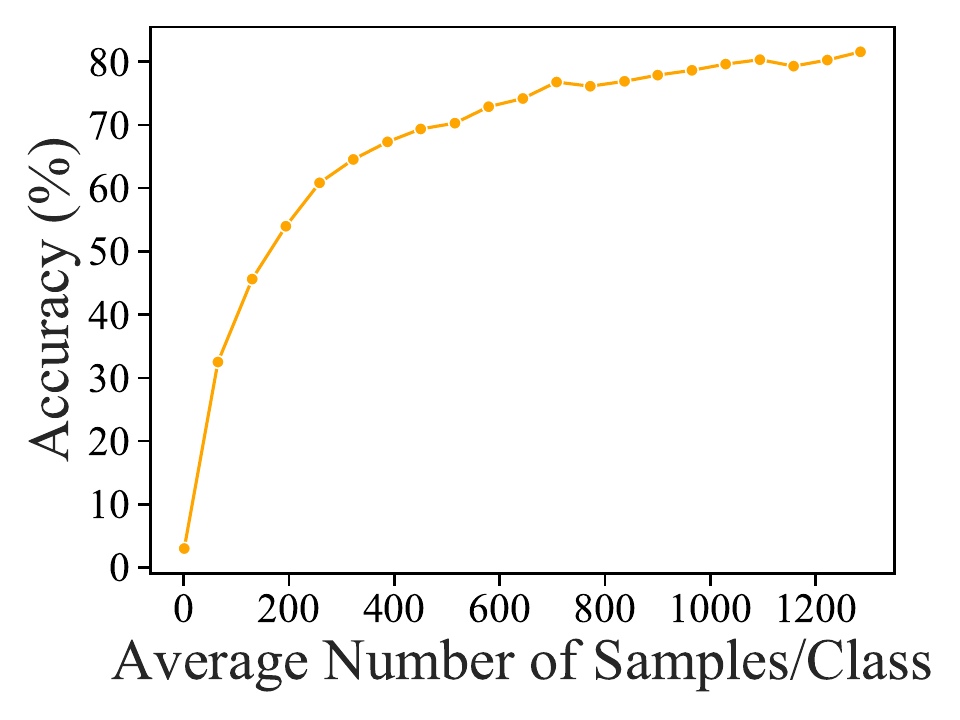}}  
\caption{\small \new{We plot the test accuracy as a function of the average number of samples per class. Observe that as the average number of samples per class increases, so does the accuracy.}} 
\label{fig:miniimagenet_frac_dataset_accuracy}
\end{figure}

\new{The results from Figure~\ref{fig:miniimagenet_frac_dataset_accuracy} suggest that as the number of samples per class goes down, so does the accuracy.  This is the case with increased sharding for complex learning tasks.}

 \end{document}